\documentclass[]{pasj01}
\draft

\begin{document} 
\Received{2016/01/26}
\Accepted{2016/11/13}

\title{Understanding the General Feature of Microvariability in {\it Kepler} Blazar W2R~1926$+$42}

\author{Mahito \textsc{Sasada}\altaffilmark{1,}\altaffilmark{2}}%
\altaffiltext{1}{Institute for Astrophysical Research, Boston University, 725 Commonwealth Avenue, Boston, MA 02215, USA}
\altaffiltext{2}{Department of Astronomy, Graduate School of Science, Kyoto University, Kitashirakawa-Oiwake-cho, Sakyo-ku, Kyoto 606-8502, Japan}
\email{sasada@bu.edu}

\author{Shin \textsc{Mineshige}\altaffilmark{2}}

\author{Shinya \textsc{Yamada}\altaffilmark{3}}
\altaffiltext{3}{Department of Physics, Tokyo Metropolitan University, Minami-Osawa 1-1 Hachioji, Tokyo 192-0397, Japan}

\author{Hitoshi \textsc{Negoro}\altaffilmark{4}}
\altaffiltext{4}{Department of Physics, Nihon University, 1-8 Kanda-Surugadai, Chiyoda-ku, Tokyo 101-8308, Japan}


\maketitle

\begin{abstract}

 We analyze the {\it Kepler} monitoring light curve of a blazar 
 W2R~1926$+$42 to examine features of microvariability by means of 
 the "shot analysis" technique. We select 195 intra-day, flare-like 
 variations (shots) for the continuous light curve of Quarter~14 with a 
 duration of 100~d. In the application of the shot analysis, an averaged 
 profile of variations is assumed to converge with a universal profile 
 which reflects a physical mechanism generating the microvariability in 
 a blazar jet, although light-variation profiles of selected shots show a 
 variety. A mean profile, which is obtained by aligning the peaks of the 
 195 shots, is composed of a spiky-shape shot component at 
 $\pm$0.1~d (with respect to the time of the peak), and two slow 
 varying components ranging from $-$0.50~d to $-$0.15~d and from 
 0.10~d to 0.45~d of the peak time. The former spiky feature is well 
 represented by an exponential rise of 0.043$\pm$0.001~d and an 
 exponential decay of 0.061$\pm$0.002~d. These timescales are 
 consistent with that corresponding to a break frequency of a power 
 spectrum density calculated from the obtained light curve. After 
 verification with the Monte-Carlo method, the exponential shape, but 
 not the observed asymmetry, of the shot component can be explained 
 by noise variation. The asymmetry is difficult to explain through a 
 geometrical effect (i.e. changes of the geometry of the emitting region), 
 but is more likely to be caused by the production and dissipation of 
 high-energy accelerated particles in the jet. 
 Additionally, durations of the detected shots show a systematic 
 variation with a dispersion caused by a statistical randomness.
 A comparison with the 
 variability of Cygnus~X-1 is also briefly discussed.

\end{abstract}

\section{Introduction} \label{Introduction}

Blazars have relativistic jets whose axes are nearly aligned to the line
of sight \citep{Blandford79,Antonucci93}. In principle, timescales of
brightness variations in blazars are related to sizes of emitting regions 
and the speeds of motions in relativistic jets. Variations, however, have a 
variety of timescales ranging from minutes to decades. The power 
spectrum density (PSD) of a blazar can be fit by a power law, 
which means that variations of blazars follow a noise-like behavior
\citep{Kataoka01}. Brightness variations of blazars could be affected 
by a variety of physical conditions: size and speed of the emission 
region, changes in magnetic field, etc. Shorter-timescale variations 
can reflect physical processes in the inner emitting regions of a jet
without any direct relation to the other, more slowly varying component(s).
The study of short-timescale fluctuations is therefore important toward 
investigation of the origin of variation in blazar jets.

Blazars show variations having a timescale of less than one day, termed 
"microvariability". Such microvariability has been reported 
over wide ranges of wavelengths from radio \citep{Quirrenbach92}, 
to optical \citep{Carini90}, X-ray \citep{Kataoka01}, and TeV
bands \citep{Aharonian07}. The {\it Fermi} space telescope scans
the entire $\gamma$-ray sky every three hours, and has detected flares,  
large-amplitude variations, in a number of blazars \citep{Abdo11}. 
\citet{Saito13} reported that a few flares in PKS~1510$-$089 exhibited asymmetric 
profiles. \citet{Nalewajko13}, however, reported that there was a great 
variety of flare shapes and duration among 40 flares that he studied, so that
the flares cannot be 
described by a simple rise and decay. It is not easy to extract detailed 
features of flare-like variations in the $\gamma$-ray band, because the time 
required to measure the $\gamma$-ray flux with {\it Fermi} and {\it AGILE} is 
usually longer than 3~hours 
(in the exceptional case of minutes-timescale $\gamma$-ray variation in 
3C~279 reported by \cite{Ackermann16}),
since the number of detected photons is limited. 
A statistical study of a sizeable number of variation events with higher 
time resolution and with good photon statistics is needed to extract the 
general features of microvariability in an effort to understand the underlying 
physics of relativistic jets. 

The blazar W2R~1926$+$42 has a synchrotron spectral energy distribution 
(SED) that peaks at a frequency of below $10^{13}$~Hz. The object is classified 
as a low-frequency peaked BL Lac object at a redshift $z=0.154$ that is 
estimated from two absorption lines in the spectrum of its host galaxy \citep{Edelson12}. 
\citet{Edelson13} also reported numerous flares on timescales as short 
as 1 day in the {\it Kepler} light curve with 30-minute time sampling in 
Quarters~11 and 12. Continuous optical monitoring of 
W2R~1926$+$42 with denser (1 minute) time sampling by {\it Kepler} 
\citep{Borucki10} in Quarter~14 detected considerable microvariability 
of the flux. 

We wish to stress here that it is of limited use to examine a variety 
of individual shapes of the time profiles of flux variations. In order 
to gain physical insight, it is more useful to examine the average 
properties. Therefore, in this paper 
we adopt a stacking analysis, so-called "shot analysis", to 
obtain a mean profile of rapid variations.
The paper is organized as follows. Details 
of the {\it Kepler} observed light curve and its PSD are described in
\S\ref{sec:Observation}. The methods of time series analysis, shot 
analysis, bootstrap method, and Monte-Carlo method, are described in
\S\ref{sec:Time}. Observational features of the mean profile of rapid 
variations and its validation by a Mote-Carlo simulation are reported in 
\S\ref{sec:Results}.
We then discuss the mechanism of variations as derived from general
features of the rapid fluctuations in \S\ref{sec:Discussion}. Several concluding 
remarks based on our results are provided in \S\ref{sec:Conclusion}.

\section{Observation and Light curve} \label{sec:Observation}
\subsection{{\it Kepler} Data} \label{sec:Kepler}

{\it Kepler} monitored over a 100,000 objects in the Cygnus 
region, obtaining continuous light curves with two timing settings, 
long (30-minute) and short (1-minute) integrations. 
W2R~1926$+$42 is listed in the {\it Kepler} target list. A continuous 
light curve with the long cadence has been obtained since Quarter~11. 
In Quarter~14, the object was monitored in the short cadence 
mode for 100~d. We have produced the calibrated "SAP\_FLUX" light curve 
with 1-minute time resolution by the automated {\it Kepler} data 
processing pipeline \citep{Jenkins10}.

\subsection{Light Curve} \label{sec:LC}

\begin{figure*}
\begin{center}
\begin{tabular}{c}
\includegraphics[width=14cm]{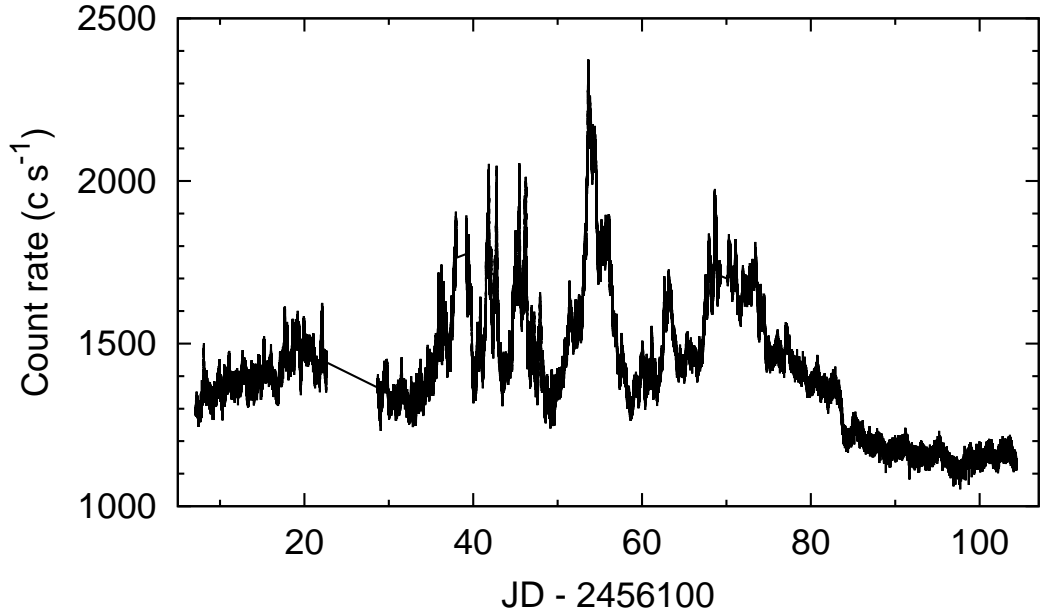}
\end{tabular}
  \caption{Light curve obtained by the {\it Kepler} spacecraft over the entire
 Quarter~14 period. The object was monitored for 100~d with 1-minute time
 resolution. }
  \label{fig:lc}
\end{center} 
\end{figure*}

Figure~\ref{fig:lc} shows an optical light curve of the object obtained by
{\it Kepler}. The blazar displayed violent variability over various
timescales ranging from several tens of minutes to over 10 days during
this monitoring. The light curve is composed of not only 
large-amplitude, long-term variations such as that ranging from
JD~2456150 to 2456160, but also numerous flare-like variations with
timescales $<1$~d. These rapid variations exist throughout this
entire monitoring period. 
A variety of profiles is apparent in these rapid variations. 
Figure~\ref{fig:lcElect} shows all profiles of the detected rapid variations.

\subsection{Power Spectrum Density} \label{sec:PSD}

Power spectrum density (PSD) analysis is one of the best ways to quantify 
time-series data. We calculate the PSD of the {\it Kepler} light curve to explore 
whether there is a characteristic timescale or not. We separate the observed light curve into five
epochs, each with a duration of approximately 20~d, and calculate PSDs at each 
epoch. In doing so, we implicitly assume that the PSD is stationary throughout 
the entire range of the light curve. We average 20 continuos power estimates
and calculate the standard error on a logarithmic scale \citep{Papadakis93}.
The standard error would contain the systematic one affected by the variation
of PSDs in each epoch.

\begin{figure}
\begin{center}
\begin{tabular}{c}
\includegraphics[scale=0.9]{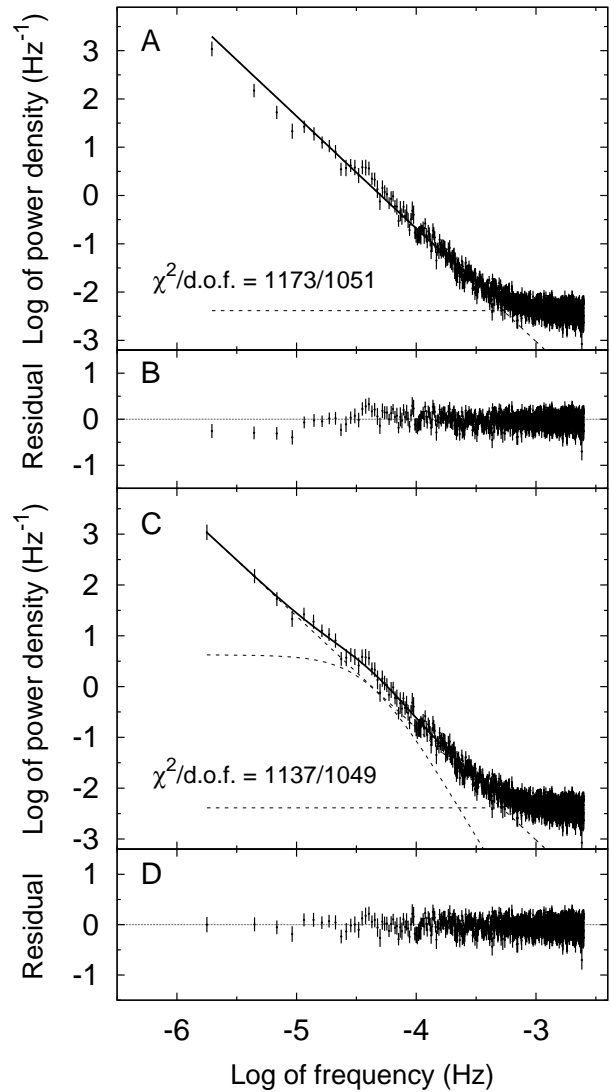}
\end{tabular}
  \caption{ Power spectrum densities calculated from the observed
    light curve. Panels (A) and (B) show the observed PSD, the
    best-fit power-law function and constant, and its residuals. Panels (C) and
    (D) show the PSD, the best-fit function of (\ref{eq:Lor}),
    and its residuals. Dashed lines in panel (A) and (C) show individual 
    components of the best-fit functions. Goodnesses of fit, 
    ${\chi}^{2}$ per degrees of freedom (d.o.f), are also indicated.}
  \label{fig:psd}
\end{center} 
\end{figure}

Figure~\ref{fig:psd} shows the PSDs calculated from the observed light curve. 
We calculate the best-fit power-law function, adding a constant value in 
panel (A), by evaluating the goodness-of-fit with a ${\chi}^{2}$ test; 
\begin{eqnarray}
  {\chi}^{2}=\sum\left(\frac{P_{\rm obs}-P_{\rm expect}}{{\sigma}_{\rm err}}\right)^{2}, \label{eq:chi}
\end{eqnarray}
where $P_{\rm obs}$ and $P_{\rm expect}$ are observed and
expected PSDs, and ${\sigma}_{\rm err}$ is the uncertainty in the
observed PSD. Residuals in panel (B) indicate a discrepancy at lower 
frequencies. We add a squared Lorentzian component to the power-law 
function as; 
\begin{eqnarray}
  P_{\rm model}(f)=A\;f^{-\alpha} + \frac{B}{[1+(f/f_{\rm br})^{2}]^{2}} + C,  \label{eq:Lor}
\end{eqnarray}
where $\alpha$ is a spectral index and $f_{\rm br}$ is a break frequency. 
The best-fit function is shown in panel (C). Dashed lines represent 
individual terms of function~(\ref{eq:Lor}). The goodness of fit for the
best-fit function improves over that for the best-fit power law. 
The existence of this curvature has already been pointed out
\citep{Edelson13,Mohan16}.

The curvature in the observed PSD indicates that the light 
curve has the characteristic timescale related to the break frequency of the 
PSD. The break frequency of the best-fit function~(\ref{eq:Lor}) is 
4.1$^{+0.6}_{-0.5}{\times}10^{-5}$~Hz, corresponding to 
$f_{\rm br}=(2{\pi}{\tau}_{\rm br})^{-1}$, with
${\tau}_{\rm br} = 0.045\pm0.006$~d, for 
time-symmetry exponential shots \citep{Negoro01}. 

If the variation is caused by flickering or 1/f fluctuation, there 
is no physical significance in the rapid variation associated with the 
higher-frequency PSD component, because a higher-frequency variation 
is generated as a result of a longer-timescale fluctuation. This fluctuation, 
however, should not have a characteristic timescale. The 
observed characteristic timescale indicates the underlying physics
associated with that timescale.

\section{Time Series Analysis Techniques} \label{sec:Time}

\subsection{Shot Analysis} \label{sec:shot}

Frequency-domain (e.g. PSD) analyses are easy to perform, but are
difficult to relate to physical mechanisms. One would prefer
a time-domain study that is useful for investigating the 
physical mechanisms of flare-like variations. This is, however, not easy 
to accomplish, especially when the photon statistics are insufficient, 
because it is difficult to perform detailed studies on a section of 
the observed data with high observational uncertainties. Additionally, 
observed flares in blazars usually have a variety of shapes 
\citep{Nalewajko13}. Thus, it is difficult to extract common features of 
blazar flare-like variations by studying only individual events.

Shot analysis proposed by \citet{Negoro94} is one of the best way to 
study the common features of variation. The shot analysis calculate the
mean of flare-like variation events by stacking these events. This mean
can be reduced the influence of the variety of shapes at individual events 
as well as the variations caused by the observational uncertainty. 
Here, the averaged profile of flare-like variations is assumed to 
converge with a universal one. In other words, if samples of flare-like 
variations are innumerable, its average profile should be coincident with 
the universal one reflecting general features of flare-like variations. 
If the variability has the characteristic timescale, the phase information
can be extracted from the mean profile of the variations associated with 
the characteristic timescale, regardless of whether the variability ascribes 
continuous or discrete processes. Hence we can investigate the physical 
mechanism associated with the characteristic-timescale variation from the 
mean of flare-like variation events.

We apply the shot analysis to the light
curve of W2R~1926$+$42 obtained by {\it Kepler} in order to generate a
mean profile of flare-like rapid variations and to study its
general features without being distracted by features specific to 
individual events. We adopt the following procedures to select rapid 
variations.
\begin{enumerate}
  \item Estimate the observational uncertainty in the light curve 
  \item Select rapid variations from the light curve as representatives 
           of flaring events 
  \item Approximate a long-term slow-varying component 
           with a polynomial function in each candidate
  \item After subtracting the long-term components, select the rapid 
           variations with variation amplitudes that are four times larger 
           than the observational uncertainty
\end{enumerate}
We define these representative variations as shots. After identifying 
the shots, we stack them aligning their peaks, and calculate the mean 
profile. An example of shot detection is shown in 
\S\ref{sec:example}.

The observational uncertainty is estimated as follows.
There are two possibilities for varying the observed brightness: 
intrinsic variation of the object and variation from observational
uncertainty. The latter can be dominant over the former within a short 
time period. We calculate differences of fluxes between two neighboring 
points as, ${\Delta}F(t_{n})=F(t_{n+1})-F(t_{n})$, 
where $F(t_{n+1})$ and $F(t_{n})$ are the $(n+1)$-th and $n$-th photon 
fluxes at those times. We define the standard deviation $\sigma$ of 
${\Delta}F(t_{n})$ as the observational uncertainty, $\sigma=$17.15 
count~s$^{-1}$. At this time, we do not include values of ${\Delta}F(t_{n})$ 
with long time differences ($>2$~min) to avoid contaminations of data 
that might have components with large intrinsic variations
because the light curve often has long blank periods caused 
by instrumental limitations.

Rapid variations are often superposed on long-term variations in
the light curves of blazars \citep{Sasada08}. We approximate the 
long-term baseline component by a second-order local polynomial that fits 
the trend of the light curve when the rapid fluctuations are ignored. 
We identify a
shot when the estimated amplitude of the rapid variation, without the 
contribution of the baseline component, is larger than our threshold criterion,
$>$4$\sigma$.

The peak time of the shot is defined as the time of maximum flux in 
the light curve after subtraction of the baseline component. We calculate 
the mean profile of detected shots by stacking numerous shots after 
aligning their peaks. Here, we average the shots without 
subtracting the baseline components. The final shape of the mean 
profile does not depend on the existence of the baseline component in 
each shot, since the baseline component of the mean profile 
should be smoothed and asymptotically close to constant in time.

\subsection{Non-parametric Bootstrap Method} \label{sec:boot}

The bootstrap method that we employ determines the distribution of an estimator
or test statistic by resampling either the data or a model derived from the
data \citep{Efron79,Efron94}. The bootstrap method first provides an
approximation to the probability distribution of the estimator 
in the detected samples. Then, the
coverage probabilities of confidence intervals can be estimated from the
probabilities of the distributions. We apply a non-parametric bootstrap
method to the dataset of shots, and try to estimate a systematic
uncertainty and a confidence interval of the mean profile of shots,
as well as confidence intervals of best-fit parameters of 
the function that reproduce the mean profile of resampled shots. 

We identify 195 shots from the {\it Kepler} monitoring light curve 
of W2R~1926$+$42, as presented below in \S\ref{sec:Shot}.
We resample these shots to produce 195 resampled shots. We then calculate 
the mean profile using these resampled shots
(hereafter, resampled mean profile).
We produce 10$^{4}$ resampled mean profiles with different resamplings 
by following this procedure. 

The best-fit parameters of the function representing the shots can be 
calculated from each resampled mean profile. Confidence intervals of parameters 
of the function can be evaluated from the probability distributions of 
the best-fit parameters calculated from each resampled mean profile. 

\subsection{Monte-Carlo Simulation of Noise Variation} \label{sec:pseudo}

Time-series analyses in past studies have found that blazar 
variations are similar to aperiodic red noise, meaning that variations
on longer timescales have greater power. We evaluate the 
characteristics of noise process by applying the above shot 
analysis. We then compare the characteristics with the 
observed features of the mean profile of shots, and examine the 
difference between the shot features estimated from the noise 
process and the observed light curve.

We adopt a Monte-Carlo method that applies the inverse Fourier
transform of the observed PSD to represent aperiodic noise-like 
linear time series with a additive sine model \citep{Timmer95}. At 
this point, we assume the best-fit model of function~(\ref{eq:Lor}) 
when the observed PSD is used as the input. Furthermore, 
we adopt a generation process of non-linear time series with a 
multiplicative sine model proposed by \citet{Uttley05}. We actually 
calculate a simulated linear time series by a fast Fourier transform 
technique, with the PSD including random fluctuations. We then 
convert this to a simulated non-linear time series through an exponential 
transform.

The time resolution of a simulated variation is 60~s, and its duration
is 100~d (corresponding to the estimations of 144000 points). The 
mean and standard deviation, ${\mu}_{l}$ and ${\sigma}_{l}$, of the generated 
simulated linear variation, $l(t)$, is adjusted to that of the observed 
light curve (equal to 1413.7 count~s$^{-1}$ and 197.3 count~s$^{-1}$). 
To calculate a simulated non-linear variation, $x(t)$, first the generated 
simulated linear variation is offset to ${\mu}_{l}$=0. Next, its variation is 
converted through an exponential transform. Then the mean and standard 
deviation of its non-linear variation, ${\mu}_{x}$ and ${\sigma}_{x}$, are 
represented as 
\begin{eqnarray}
  {\mu}_{x}={\rm exp}\left[\frac{1}{2}{\sigma}_{l}^{2}\right], \label{eq:mux}  \\
  {\sigma}_{x}={\rm exp}\left[{\sigma}_{l}^{2}\right]\left( {\rm exp}\left[{\sigma}_{l}^{2}\right] - 1 \right). \label{eq:six}
\end{eqnarray}
A fractional rms, ${\sigma}_{\rm frac}$, is defined as the standard deviation 
divided by the mean, which corresponds to a skewness of $x(t)$. This 
fractional rms is characterized by equations~(\ref{eq:mux}) and 
(\ref{eq:six}) as,
\begin{eqnarray}
  {\sigma}_{\rm frac}=\sqrt{{\rm exp}\left[{\sigma}_{l}^{2}\right]-1}.
\end{eqnarray}
In this simulation, we assume that ${\sigma}_{\rm frac}$ of the 
simulated non-linear variation is equal to 0.64. Finally, the mean 
and standard deviation of $x(t)$ are adjusted to the observed 
values.

We select local peaks of the simulated linear variation. Mean 
profiles of local peaks in the simulated linear and non-linear 
variations are calculated from 195 peaks, the 
same as the number of detected shots (see at \S\ref{sec:mean} below).

\section{RESULTS} \label{sec:Results}

\subsection{Shot Analysis} \label{sec:Shot}

The PSD analysis reveals that the variability of W2R~1926$+$42 has 
the characteristic timescale as the curvature of the PSD. This indicates 
the physical background associated with the variation timescale. We 
perform the shot analysis to the {\it Kepler} light curve to extract phase 
information of variation with the characteristic timescale of the object. 

A large number of hour-scale episodes of rapid variations 
are detected in the light curve. We generate the mean profile of shots that 
satisfy the definition mentioned at \S\ref{sec:shot} to extract 
general features of the rapid variation.

\subsubsection{Example of Shot Detection} \label{sec:example}

\begin{figure}
\begin{center}
\begin{tabular}{c}
\includegraphics[scale=0.9]{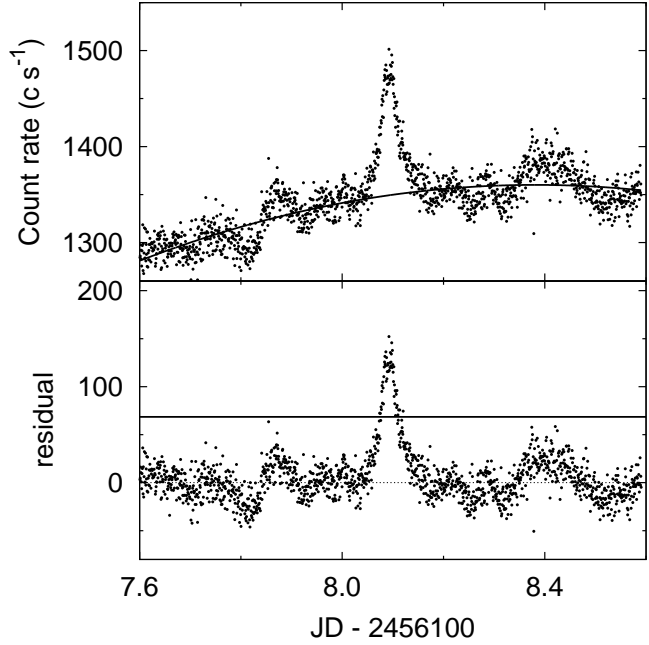}
\end{tabular}
  \caption{Example of shot detection. Top panel shows a light curve with 
  a detected shot and a polynomial function approximating a long-term 
  baseline component underlying the shot shown as solid line. Bottom panel shows the light 
  curve after subtraction of the best-fit polynomial function. Solid line 
  indicates the threshold for detecting shots.}
  \label{fig:exam}
\end{center} 
\end{figure}

Many temporal surges in flux are seen in the light curve.
We select these surges and discriminate against large-amplitude variations 
according to the criteria given in \S\ref{sec:shot}.
Figure~\ref{fig:exam} shows an example of shot detection. Only the peak at 
JD~2456108.09 can be identified as a shot. We estimate the 
amplitude of the variation by subtracting the best-fit polynomial 
baseline component of the light curve. Small-amplitude surges seen
on JD~2456107.86 and 2456108.40 do not satisfy the criteria of shots. 
We detect 195 shots from the entire light curve. 
All shots are displayed in Figure~\ref{fig:lcElect}.

\subsubsection{Mean Profile of Shots} \label{sec:mean}

\begin{figure}
\begin{center}
\begin{tabular}{c}
\includegraphics[scale=0.7]{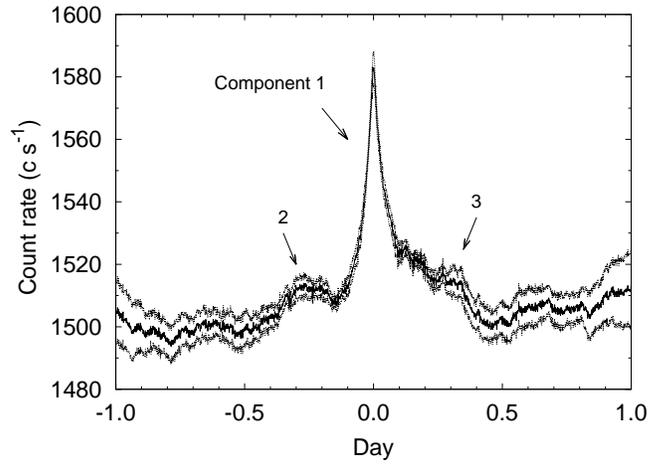}
\end{tabular}
  \caption{ The mean profile of detected shots. Dotted lines show standard 
  deviations of the mean profile. These standard deviations are calculated 
  from mean profiles of shots detected at light curves which are separated 
  six epochs. See the text for details.}
  \label{fig:shot}
\end{center} 
\end{figure}

We calculate the mean profile of 195 detected shots. Figure~\ref{fig:shot} 
shows the mean profile of shots.
The plot omits the count rate at the peak time ($t=0$), since positive
fluctuations of the counts at this time are summed systematically
\citep{Negoro94}. 

There are three mainly components at the mean profile: 
a spike-like component in $\pm$0.1~d (component~1), and slowly varying 
components ranging from $-$0.50 to $-$0.15~d and from 0.10 to 0.45~d 
(component~2 and 3, respectively). The increase and decrease of 
component~1 are well reproduced by an exponential rise and decay. 
The peak is spiky but smoothly connected from the rise to decay phases. 

There is a possibility that shots evolve with time. The systematic 
uncertainty of the mean profile of shots should include the influence of time 
evolution of shots. To estimate the systematic uncertainty, we separate 
the light curve into 6 epochs; (1) JD~2456106--2456125, (2) 
2456128.6--2456143.6, (3) 2456143.6--2456158.6, (4)
2456158.6--2456173.6, (5) 2456173.6--2456188.6, and (6)
2456188.6--2456205. Individual mean profiles are calculated 
from shots located in each epoch. We estimate variances of the mean 
profiles in each time step. We then calculate the standard deviation of the 
weighted mean, because the numbers of shots in each epoch are 
different. Here we normalize fluxes of individual mean profiles within 
$\pm$1~d to average fluxes over all of the mean profiles, since 
components~2 and 3 are distributed within $\pm$1~d.

Figure~\ref{fig:shot} shows the mean profile of shots and calculated 
standard deviations of the weighted means at each time step. 
Amplitudes of components~2 and 3 are larger than the
standard deviations. Therefore, components~1, 2 and 3 of 
the mean profile of shots can be regarded as real phenomena, not
artifacts of the systematic uncertainty of the sampling of shots.

\subsubsection{Model for Component 1 of the Mean Profile} \label{sec:model} 

Component~1 is distributed around the peak time of mean profile of shots.
General features of shots can be extracted from this component.
First, we characterize the shape of component~1 to a function proposed 
by \citet{Abdo10}:
\begin{eqnarray}
  F(t) = F_{0}\;[e^{-t/T'_{\rm r}}+e^{t/T'_{\rm d}}]^{-1} + F_{\rm c}, \label{eq:con}
\end{eqnarray}
where $T'_{\rm r}$ and $T'_{\rm d}$ are variation timescales of the rise
and decay phases, $F_{\rm c}$ represents a constant level underlying
component~1, and $F_{0}$ measures the amplitude of the shot. 

We evaluate its goodness-of-fit of this function with a ${\chi}^{2}$ test. 
The standard deviation of the weighted mean between six mean profiles of 
shots detected from individual epochs is adopted as an estimate of the 
systematic error, ${\sigma}_{\rm err}$, to calculate the ${\chi}^{2}$ of component~1. 
The goodness of fit of component~1 by function~(\ref{eq:con}) is 
${\chi}^{2}=$218.7.

\begin{figure}
\begin{center}
\begin{tabular}{c}
\includegraphics[scale=0.9]{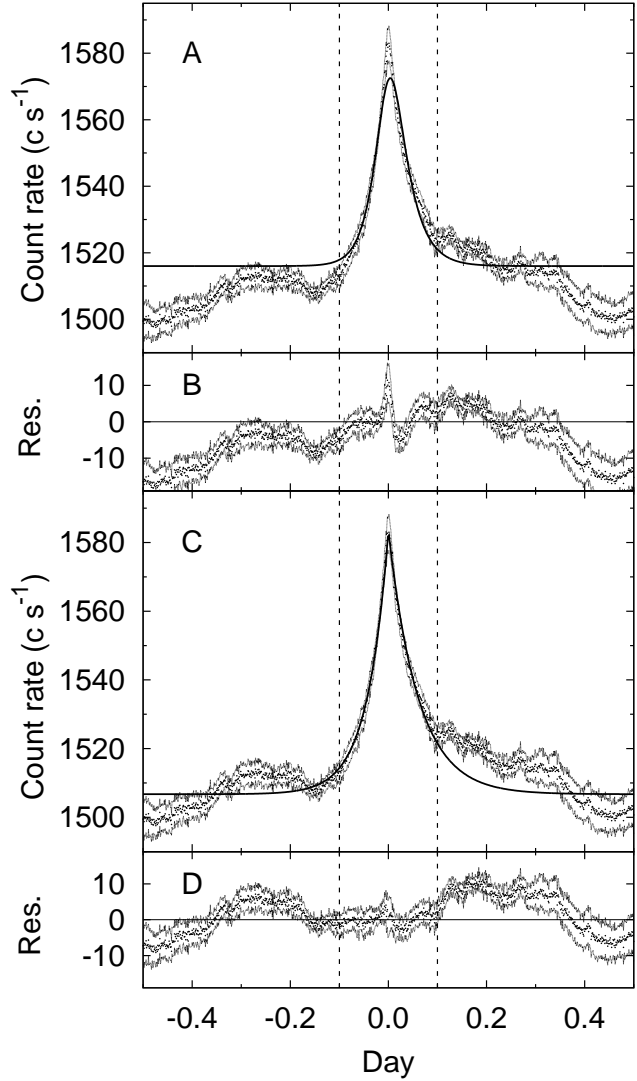}
\end{tabular}
  \caption{Best-fit functions superposed on the mean profiles of
 shots. Panels (A) and (C) show the mean profile of shots and the
 best-fit functions~(\ref{eq:con}) and (\ref{eq:exp}). Panels (B) and
 (D) show residuals between the mean profiles and the estimated
 best-fit functions. Dashed lines show the fitted range.}
  \label{fig:fit}
\end{center} 
\end{figure}

In comparison, we apply another function (cf. Negoro et~al. 1994):
\begin{eqnarray}
  F(t) = \left\{ 
  \begin{array}{ll}
    F_{0}\; e^{|t|/T_{\rm r}} + F_{\rm c}; & (t < 0) \\
    F_{0}\; e^{-|t|/T_{\rm d}} + F_{\rm c}; & (t > 0), \label{eq:exp}
  \end{array}
  \right.
\end{eqnarray}
where $T_{\rm r}$ and $T_{\rm d}$ are e-folding times of the rise and decay,
and $F_{\rm c}$ and $F_{0}$ are the same in the case of
function~(\ref{eq:con}). 
Since the number of degrees of freedom is equal to that 
in the case of function~(\ref{eq:con}), we can compare values of ${\chi}^{2}$ 
between the best-fit functions of (\ref{eq:con}) and (\ref{eq:exp}) directly. 
The goodness of fit of function~(\ref{eq:exp}) is $\chi^2$=52.6, which 
is better than that of function~(\ref{eq:con}). Figure~\ref{fig:fit} shows the 
applied functions with the best-fit parameters superposed on the 
mean profile and its residuals. Although function~(\ref{eq:con}) 
shows obvious residuals during the peak time, as shown in panels (A)
and (B) of figure~\ref{fig:fit}, the residuals in the case of 
function~(\ref{eq:exp}) are suppressed, as shown in panels (C) and 
(D). This indicates that the mean profile is more spiky than expected by 
function~(\ref{eq:con}). Thus, function (\ref{eq:exp}) is better suited than 
function~(\ref{eq:con}) to represent the component~1 of the mean profile 
of shots.

The best-fit parameters of function~(\ref{eq:exp}) are shown in 
table~\ref{tab:exp}. The best-fit e-folding times of the rise and decay 
phases, at 0.043~d and 0.061~d, are different.
We note that the average of these timescales, 0.052$\pm$0.003~d, is 
consistent with the variation timescale 
calculated from the break frequency of the PSD within 1-sigma confidence 
level (\S\ref{sec:PSD}). This result indicates that the component~1 
corresponds to the curved feature seen in the observed PSD. 

Apparently, the rise timescale of the shot in figure~\ref{fig:fit} is
slightly shorter than that of the decay. Is this difference statistically
significant? To answer to this question, we calculate confidence
intervals of the parameters of function~(\ref{eq:exp}) and the ratio 
between the rise and decay e-folding times.

We estimate the rise and decay timescales from the best-fit 
functions~(\ref{eq:exp}) of six mean profiles of shots selected
from different epochs. The rise and decay timescales are 
different in epochs; 0.077 and 0.124 (epoch 1), 0.027 and 0.041 
(epoch 2), 0.029 and 0.054 (epoch 3), 0.077 and 0.127 (epoch 4), 
0.091 and 0.090 (epoch 5), and 0.094 and 0.236 (epoch 6), 
respectively. These timescales lead that 
the characteristic timescale is variable in time. We calculate 
ratios of the rise to decay timescales to understand how 
asymmetric the mean profiles are. Most of ratios are less than 1, 
and only one profile is approximately equal to unity. These 
ratios are distributed from 0.40 to 1.01. We calculate the 
weighted mean of ratios of timescales and its standard deviation
associated with the numbers of shots as 0.63$\pm$0.11. This 
indicates that the profiles of shots highly tend to show a fast-rise 
and slow-decay feature.

\begin{figure}
\begin{center}
\begin{tabular}{c}
\includegraphics[scale=0.9]{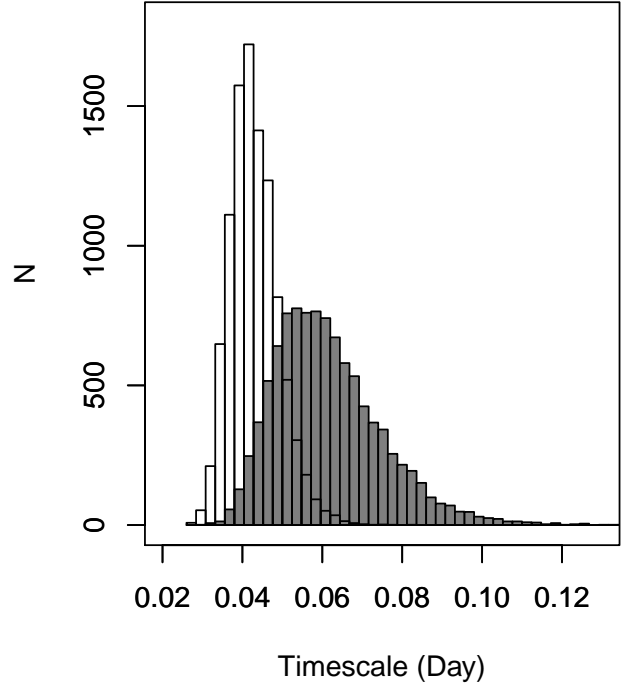}
\end{tabular}
  \caption{Histograms of $T_{\rm r}$ (white) and $T_{\rm d}$ (gray)
 of the best-fit function~(\ref{eq:exp}). The best-fit parameters 
 are calculated from 10$^{4}$ resampled mean profiles generated by the 
 non-parametric bootstrap approach. See text for details.}
  \label{fig:Br}
\end{center} 
\end{figure}

The mean profile of shots may have a deviation associated with
the variation of individual shapes of 195 shots. To verify the effect of 
this deviation to the asymmetry of the mean profile, we evaluate the 
difference between the rise and decay e-folding times by using the 
non-parametric bootstrap method as mentioned in \S\ref{sec:boot}. 
We generate 10$^{4}$ resampled mean profiles, and calculate the 
best-fit parameters of function~(\ref{eq:exp}) for component~1 of 
each resampled mean profile. The confidence levels of parameters 
can be estimated from distributions of best-fit parameters. 
Figure~\ref{fig:Br} shows distributions of $T_{\rm r}$ and
$T_{\rm d}$. This clearly shows that these parameters are 
differently distributed. 

We apply the Wilcoxon rank-sum test 
which is a non-parametric significance test (also referred to as the
Mann-Whitney U-test), to the distributions of $T_{\rm r}$ and 
$T_{\rm d}$ \citep{Wilcoxon45,Mann47}. Since the p value of the 
Wilcoxon rank-sum test is less than $10^{-15}$, we confirm that 
median values of distributions of $T_{\rm r}$ and $T_{\rm d}$ 
calculated from the detected shots are clearly different.
For these evaluations, component~1 in the mean profile 
is asymmetric in this case of 195 detected shot samples.

\begin{table}
  \tbl{Parameters of best-fit function~(\ref{eq:exp}) to
 component~1 of the mean profile of shots \label{table:fit}}{%
  \begin{tabular}{ll}
      \hline
       & Best-fit value \\ 
      \hline
      $T_{\rm r}$ (d) & 0.043$\pm$0.001 \\
      $T_{\rm d}$ (d) & 0.061$\pm$0.002 \\
      $F_{0}$ (count~s$^{-1}$) & 76.7$\pm$0.6 \\
      $F_{\rm c}$ (count~s$^{-1}$) & 1506.7$\pm$0.8 \\
      \hline
    \end{tabular}}\label{tab:exp}
\end{table}

\subsubsection{Amplitude Dependence of Mean Profiles}

\begin{figure}
\begin{center}
\begin{tabular}{c}
\includegraphics[scale=0.9]{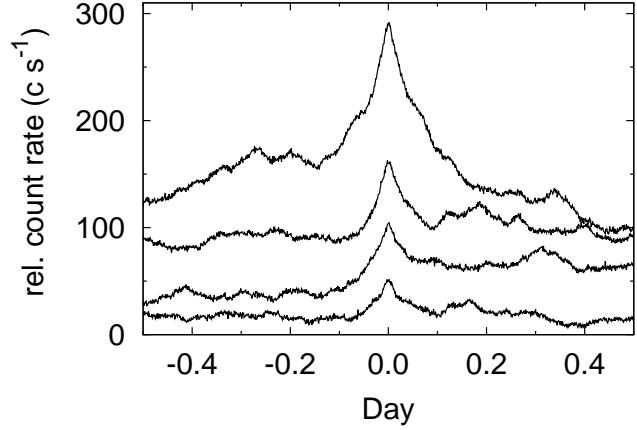}
\end{tabular}
  \caption{Mean profiles of shots with different amplitudes. From bottom
 to top, profiles are calculated from shots with amplitudes of
 4--5.7, 5.7--7, 7--9.5, and over 9.5$\sigma$, respectively. Each
 profile is offset for clarity. The vertical axis shows the relative count
 rate.}
  \label{fig:sep}
\end{center} 
\end{figure}

Since rapid variations observed in the light curve have various 
amplitudes, the shot detection is defined with a threshold of 
$>$4$\sigma$. If the profile of rapid variations depends on 
its amplitude, the calculated mean profile of shots does not reflect 
general features of the rapid variations. We separate detected shots into 
four groups based on amplitudes of 4--5.7$\sigma$, 
5.7--7$\sigma$, 7--9.5$\sigma$, and over 9.5$\sigma$, and verify 
the amplitude dependence of the shot profile by comparing the
profiles of four groups. Figure~\ref{fig:sep} shows mean profiles of 
shots with different amplitudes. All profiles have the component~1 
within $\pm$0.1~d of the center. These shot profiles do not have any
clear trend associated with their amplitudes, except for the amplitudes of 
component~1, although these profiles have local features that
are caused by the limited number of shot samples. All ratios of rise to
decay timescales in each profile are less than unity (average and standard 
deviation of these ratios are 0.69$\pm$0.18). These results
indicate that the mean profile of 195 shots reflects the general
nature of the rapid variations, and the asymmetric feature of the mean 
profile of shots is not artificial one caused by the amplitude dependence of 
shots.

\subsection{Shot Durations} \label{sec:fwhm}

Detected shots displayed a variety of shapes. It is, however, assumed that the 
averaged shape of shots converges to the exclusive shape in the shot analysis. 
If the averaged shape of shots converges to the unique shape which reflects 
the general feature of shots, durations of shots would distribute around its 
characteristic time. To validate the convergence of shot shapes, we calculate
widths of e-folding rise and decay times in each shot as durations, and 
investigate its distribution.

We calculate the shot durations as follows: First, the baseline
component under a shot is approximated by the second-order polynomial 
function. Here we set a fitting region for this approximation in each shot. 
Second, the approximated baseline component is subtracted from the light 
curve to estimate an amplitude of the shot. Thirds, e-folding timescales 
in a rise and decay phases are calculated from the subtracted light curve.
Finally, these timescales are summed as a duration of the shot.
In some cases of shots, other variation components are contaminated to
the shot components. The e-folding timescales should be longer for the
contamination. Then, we extrapolate by linear regression to expect the 
buried shot component and estimate the e-folding time from the 
expectation. 

\subsubsection{Distribution of Durations} \label{sec:distribute}

\begin{figure}
\begin{center}
\begin{tabular}{c}
\includegraphics[scale=0.9]{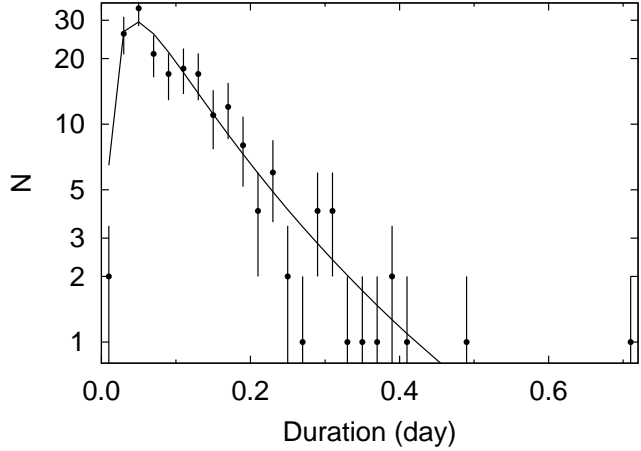}
\end{tabular}
  \caption{Distribution of estimated shot durations. Solid line shows the best-fit
  log-normal distribution.}
  \label{fig:distribute}
\end{center} 
\end{figure}

Figure~\ref{fig:distribute} shows a distribution of shot e-folding durations. Estimated 
durations of shots are ranged from 0.018 to 0.71~d, with a mean of 0.122 and a 
standard deviation of 0.097~d. In table~\ref{table:ShotElect}, there are the number, 
peak date, amplitude, duration, and rise and decay times in each shot.
The durations are distributed with a characteristic
time, and its distribution can be represented by the log-normal function:
\begin{eqnarray}
f(x)=\frac{1}{\sqrt{2{\pi}}}{\rm exp}\left[-\frac{({\rm log}x - {\rm log}{\mu})^{2}}{2{\sigma}^{2}}\right]
\end{eqnarray}
where $\mu$ and $\sigma$ represent the mean and variance of the distribution
\citep{Negoro02}. The distribution is evaluated to the log-normal function by 
the Kolmogorov and Smirnov test (with a p value of 0.53). 
The best-fit log-normal function is shown as a solid line in figure~\ref{fig:distribute}.

The mean of these durations is slightly larger than the duration (best-fit 
$T_{\rm r}$ plus $T_{\rm d}$ of 0.104~d) of the mean profile of shots. This can 
be caused by the large-side tail of the distribution of durations as evidence
that the median value of the distribution of 0.098~d is similar to the 
value of $T_{\rm r}$ plus $T_{\rm d}$ of the mean profile.  

If the durations arise following to a random manner or a provability distribution of 
a power-law function, the distribution should be a flat or no-peak shapes. If the 
distribution of durations follows to the power-law function with a lower cutoff, 
there is a physical background for the lower cutoff of this distribution, because 
the minimum duration of 0.018~d is clearly larger than the cutoff caused by the 
time-resolution limit of several times of 0.00068~d ($=$1~min). Thus, the 
duration at the peak of the distribution is brought to the physics of the jet. 
Therefore, the distribution of the durations implies that the averaged shape of shots  
is converged to the typical one which reflects the physical background of the 
object.

\subsubsection{Time Evolution of Variation Timescales} \label{sec:fwhmEv}

\begin{figure}
\begin{center}
\begin{tabular}{c}
\includegraphics[scale=0.9]{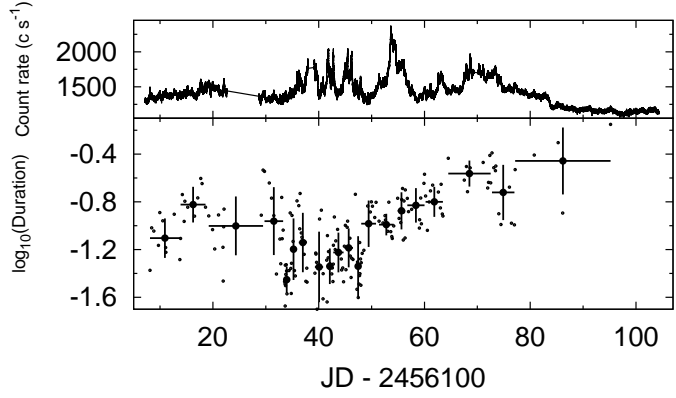}
\end{tabular}
  \caption{Time series of shot durations. The {\it Kepler} light curve (upper 
  panel) and the time variation of the durations are displayed. The time series 
  of averaged durations is also shown in the bottom panel.}
  \label{fig:duration}
\end{center} 
\end{figure}

Figure~\ref{fig:duration} shows the light curve, and time series of estimated 
durations in the detected shots together with the average and standard
deviation of each 10-points duration set in a logarithmic scale. The 
averaged durations show a systematic variation with time. To validate that 
this systematic variation is not a result of a random manner, we calculate a 
$\chi^{2}/{\rm d.o.f}$ of these averaged durations to the best-fit constant 
value in the logarithmic scale. Here, we assume that durations in each 
10-point set are randomly distributed according to a log-normal probability 
density. Therefore, the standard deviation of each 10-point duration set is 
used as the  $\sigma_{\rm err}$ in equation~\ref{eq:chi}. The calculated 
$\chi^{2}/{\rm d.o.f}$ is equal to 52.3/19, which corresponds to a p value of 
1.04$\times$10$^{-5}$. This result indicates that the averaged durations
are variable with time. Thus, the shot events are associated with each 
other, not randomly distributed.

\subsection{Validation of Shot Features by Monte-Carlo Simulation} \label{sec:monte}

In \S\ref{sec:Shot}, we found general features of rapid variation in the object
by applying shot analysis to the {\it Kepler} light curve. It is not known,
however, whether the general features result from the natures of the AGN jet 
physics or statistics. 
The observed general features extracted from the mean
profile of shots should be separated into these two categories by using the 
Monte-Carlo method. We evaluate the stochastic features of the simulated 
noise variation generated by the Monte-Carlo method (see 
\S\ref{sec:pseudo}) by applying the shot analysis, and compare these 
features with the observed ones. Then we determine the causes of the 
observed features.

\begin{figure}
\begin{center}
\begin{tabular}{c}
\includegraphics[scale=0.9]{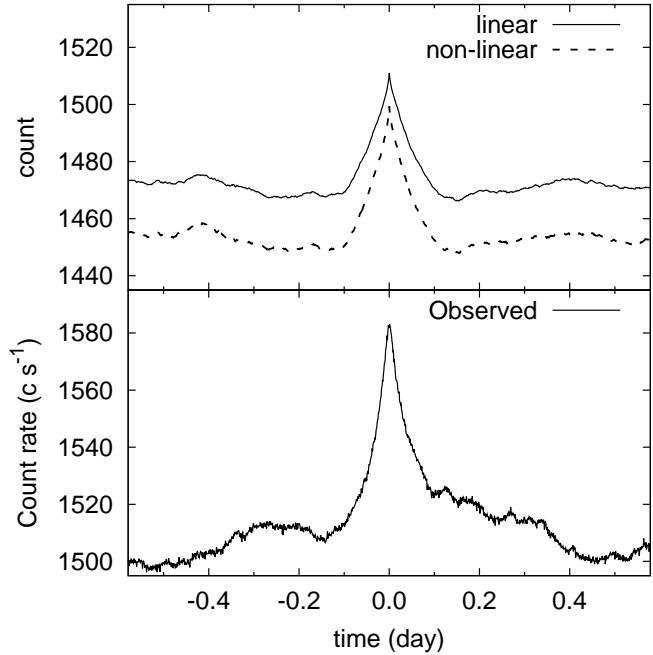}
\end{tabular}
  \caption{Mean profiles of local peaks calculated from simulated linear 
    and non-linear variations, and the observed mean profile of shots. 
    The mean profiles in the top panel are calculated from 195 local 
    peaks selected from a simulated linear (solid line) and non-linear 
    (dashed line) variations. The observed mean profile of shots in the 
    bottom panel is shown for comparison. }
  \label{fig:pseudo-mean}
\end{center} 
\end{figure}

Figure~\ref{fig:pseudo-mean} shows the mean profiles of local peaks selecting 
from simulated linear and non-linear variations, and displays the observed mean 
profile of shots for comparison. All profiles clearly have peak signals distributed 
about the origin. The rise and decay phases of the mean profiles calculated from 
both the simulated linear and non-linear variations are consistent with the 
exponential rise and decay forms, which are also consistent with the observed 
ones. Durations of the peak component in the mean profiles of local peaks in 
the linear and non-linear variations are approximately equal to the duration of 
component~1 ($\pm$0.1~d). This similarity can be caused by having the 
observed and reference PSDs having the same break frequencies, equal to 
4.1$\times$10$^{-5}$~Hz.

Averages of e-folding times in the rise and decay phases of profiles 
calculated from the simulated linear and non-linear variations are 
approximately equal to 0.084 (rise in linear), 0.084 (decay in linear)
0.078 (rise in non-linear) and 0.078~d (decay in non-linear), 
respectively. Standard deviations of rise and decay e-folding times 
calculated from 10$^{3}$ simulated linear and non-linear variations are 
almost the same value, 0.015~d. The timescales in both the linear 
and non-linear cases are longer than those of the observed mean 
profile. Furthermore, the rise and decay timescales of profiles of 
the simulated variations are almost the same. That is, the 
profiles are symmetric, while component~1 of the observed 
mean profile contains some asymmetry.

\begin{figure}
\begin{center}
\begin{tabular}{c}
\includegraphics[scale=0.8]{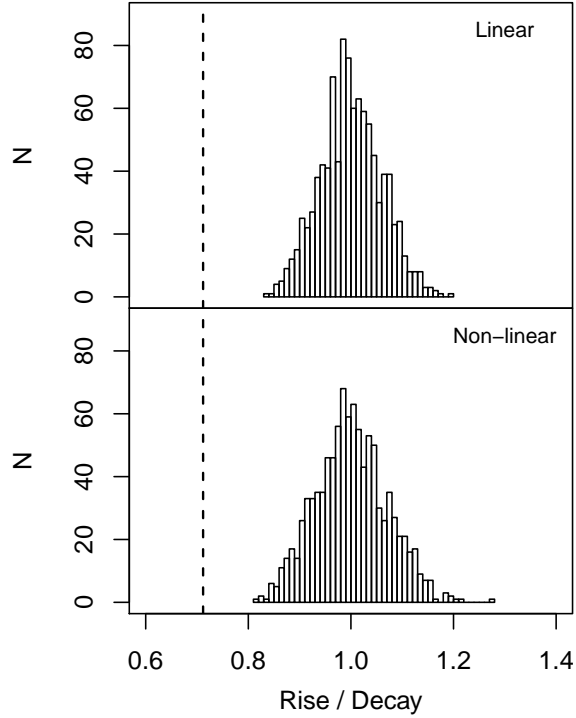}
\end{tabular}
  \caption{Histograms of 10$^{3}$ ratios between rise and decay
    timescales. The rise and decay timescales are calculated 
    as e-folding times of mean profiles of simulated linear (top panel) and 
    non-linear (bottom panel) variations. Dashed lines show
    the observed ratio of rise and decay timescales.}
  \label{fig:ratio-hist}
\end{center} 
\end{figure}

We evaluate whether the observed asymmetry can be explained by 
fluctuations in the simulated rise and decay timescales of local-peak 
profiles calculated from linear and non-linear variations. First, we 
generate 10$^{3}$ mean profiles of local peaks of simulated variations 
with the linear and non-linear processes. Second, the ratio between the 
best-fit e-folding times of the rise and decay phases is estimated 
from each mean profile. Finally, we compare the probability 
distributions of those ratios calculated from the 10$^{3}$ simulated 
linear and non-linear variations with the observed ratio between the 
e-folding times of rise and decay phases. Figure~\ref{fig:ratio-hist} 
shows histograms of 10$^{3}$ ratios between the rise and decay 
timescales of mean profiles in the cases of both the linear and 
non-linear processes. These are clearly distributed about unity. 
This result indicates that the rise and decay timescales calculated 
from linear and non-linear noise variations should be the same. 
According to the Student's $t$-test, the ratio of timescales in the 
observed mean profile ($=$0.70) is clearly different with the median 
values of ratios calculated from these simulated mean profiles (the 
p values are both less than 2.2$\times$10$^{-16}$). Thus, the 
difference between the rise and decay timescales observed in the 
mean profile of shots is caused by a physical phenomenon, not 
the result of the stochastic noise variation generated by the 
Monte-Carlo method using the PSD with a break frequency.
 
\section{Discussion} \label{sec:Discussion}
\subsection{Origin of rapid variations} \label{sec:origin}

It is poorly understood whether rapid variations of blazars are intrinsic 
phenomena or an apparent one caused by a geometrical effect, such as 
the precession of a jet axis. There are several models that explain flux
variations as apparent rather than as caused by changes in the intrinsic 
luminosity. This can occur, for example, if the Doppler factor varies
owing to changes in the viewing angle in a bent jet \citep{Villata99} 
or to the effects of gravitational lensing \citep{Chang79}. Such
models, however, predict that the averaged variation profile is approximately
symmetric in a simple geometry, because the Doppler factor or lensing should 
change symmetrically. In other words, rise and decay
timescales of an average profile of rapid variations should be roughly equal, 
whereas the estimated rise and decay
timescales are different. Thus, rapid variations can not generally be
explained by these models, but rather as intrinsic phenomena. If
particles are accelerated in the jet, the number of higher-energy particles are
increased in the emitting region during rapid variations, where they
dissipate their energies. The flux-variation profile should be asymmetric in this
particle-acceleration scenario, because the rise and decay of the
brightness are caused by different mechanisms: particle acceleration and
dissipation processes. Therefore, rapid variations are plausibly 
explained by such the particle-acceleration scenario.

\subsection{Values of the magnetic field and Doppler factor} \label{sec:mag}

The mean profile of shots calculated by the shot analysis represents the
variation with the characteristic timescale. This reflects an "averaged" 
situation of emitting regions of shots. We can estimate the common physical
values of emitting regions from the mean profile of shots.

It is plausible that the dissipation of high-energy particles during rapid
variation is caused by a synchrotron cooling. The timescale for this 
process, $\tau_{\rm syn}$, can be related to the decay
timescale $T_{\rm d}$, ${\tau}_{\rm syn}=T_{\rm d}/(1+z)$. The synchrotron 
cooling timescale is calculated as
\begin{eqnarray}
{\tau}_{\rm syn}\;{\sim}\;3.2{\times}10^{4}\;B^{-3/2}\;E^{-1/2}\;{\delta}^{-1/2}\quad{\rm s}, \label{eq:tau}
\end{eqnarray}
where $B$ is the strength of magnetic field in Gauss, $\delta$ is the Doppler
factor, and $E$ is the energy of synchrotron photon in the observer's frame in eV
\citep{Tashiro95,Sasada10}. We obtain the relation between $B$ and 
$\delta$ from the equation~(\ref{eq:tau}) and the observed $T_{\rm d}$
of the mean profile of shots as:
\begin{eqnarray}
 {\delta}=26.1\;B^{-3}\;{\left(\frac{T_{\rm d}}{0.061~{\rm d}}\right)}^{-2}, \label{eq:delB}
\end{eqnarray}
where we adopt $E$ as the {\it Kepler}-observed energy,
$E$=1.88~eV (central wavelength of the spectral response of the {\it Kepler}
instrument, $\lambda$=6600\AA; see \citep{Koch10,Van10}).
The estimated value of $T_{\rm d}$ can be regarded as typical of rapid 
variations, because the timescale is calculated from the mean profile of 
shots. Thus, the relation between $\delta$ and $B$ is typical 
of rapid variations in this object.

We calculate $\delta=209$ assuming $B$ as 0.5~G which is typical value
of $\gamma$-ray detected BL~Lac objects \citep{Ghisellini10}. The
average Lorentz factor among such BL~Lac objects is 
estimated as $\Gamma=6.1$, and most are viewed within 10$^{\circ}$ 
of the jet axis \citep{Ajello14}. The Doppler factor $\delta$ is represented 
by the Lorentz factor and the viewing angle $\theta$:
$\delta=(\Gamma-\sqrt{{\Gamma}^{2}-1}\;{\rm cos}\;{\theta})^{-1}=5.7$, 
calculated for $\theta=10^{\circ}$. The Doppler factor calculated from the
observed timescale is much larger than the typical of $\gamma$-ray detected
BL~Lac objects. If the Doppler factor is consistent with the typical, equal to 10, 
the magnetic field is approximately equal to 1.4~G, which is higher than the 
typical value of $\gamma$-ray detected BL~Lac objects. 

An alternative idea is that the high-energy accelerated particles are dissipated 
by escaping from the emitting region. The dissipation timescale corresponds 
to the size of the emitting region, $R$ and the Doppler factor, 
$R\;{\leq}\;{\delta}cT_{\rm d}$. 
The average size of emission region should be smaller than 
1.6$\times$10$^{15}$~cm calculated from the decay timescale of the mean
profile of shots. This limitation of the size of emission region is much smaller 
than that using in the multi-wavelength spectral study, roughly equal to 
10$^{17}$~cm.

\subsection{Shape of the profile of rapid variations} \label{sec:shape}

Particle acceleration and energy loss processes can lead to an asymmetric 
profile of rapid variations. There are several proposed 
particle acceleration mechanisms in blazar jets, for
example, the shock-in-jet scenario \citep{Marscher85}, and magnetic 
reconnections \citep{Giannios09}. The mean
profile reflects the general features of variations without local features 
associated with different physical situations at individual 
variations. Here we compare component~1 with a simulated time evolution of 
emission proposed by past papers.

A numerical approach for reconstructing an episode of blazar variability 
has been attempted to reproduce observed synchrotron light curves and
multi-wavelength spectra \citep{Kirk98}. \citet{Spada01} suggest that
the observed variability can be explained via the inverse Compton process
within the internal shock scenario. They proceed by 
simulating the birth, propagation and collision of shells, calculating
the spectrum produced in each collision, and summing the locally
produced spectra.

\citet{Kirk98} calculate a simulated light curve of blazar variation
assuming a model in which particles are accelerated at a shock front
and cool by synchrotron radiation in the homogeneous magnetic field. In
this model, the increase of the injection rate of accelerated high-energy
particles is constant in time. 
The flux of synchrotron radiation from the shocked region varies 
depending on the balance between the acceleration and dissipation rates 
of high-energy electrons. The observed sharp peak of the mean profile of 
shots indicates that the dominant rate changes dramatically at the peak.
Simulated light curves at the maximum
frequency of the synchrotron SED ($10^{18}$~Hz) and at a lower 
frequency ($10^{16}$~Hz) are different. The peak shape of the simulated 
light curve at $10^{16}$~Hz is more spiky than that in $10^{18}$~Hz. 
Component~1 in the observed mean profile is similar to the case of 
$10^{16}$~Hz. 
Based on the model in \citet{Kirk98}, the sharply peak of the mean 
profile of shots
implies that the synchrotron-peak frequency of the rapid variation can be 
higher than the {\it Kepler}-observed wavelength. The observed peak 
frequency of synchrotron SED is, however, below $10^{13}$~Hz
\citep{Edelson13}. Therefore, higher-energy electrons, which emit at
higher frequencies of the synchrotron spectrum, could be generated 
during the rapid variation, whereas electrons emitting the more stable 
synchrotron spectrum are produced less sporadically.

Several authors assume that the injection rate is a 
function of the Lorentz factor of the accelerated particles
\citep{Kusunose00,Bottcher10}. 
The simulated flux with a constant injection rate of high-energy
particles rises rapidly at the beginning of the
injection, after which it rises more slowly. This feature is caused by a
balance between the injection and dissipation rates of accelerated
high-energy particles. The rising phase of the observed mean profile,
however, follows a simple exponential increase. Therefore, the observed 
light curve in the rising phase does not correspond to that expected
for a constant injection rate.

\subsection{Implication for Systematic Variation of Shot Durations} \label{sec:ImpliDura}

The durations of detected shots changed both randomly and systematically 
through time. If the flux variation is caused by the high-energy electron 
acceleration to the relativistic speed, the obtained systematic change of 
shot durations implies that these acceleration events are correlated with 
each other. Several models are proposed for the mechanisms of particle 
acceleration in blazar jets, for example, the shock-in-jet scenario 
\citep{Marscher85}, and magnetic reconnections \citep{Giannios09}. The 
particle acceleration event is, however, expected to happen randomly 
based on simple situations of proposed models. In the shock-in-jet model, 
the particle acceleration is provided by a shock wave arising from a collision 
of two dense plasmas. Similarly, shocks and particle accelerations can occur 
in a field with two magnetic field lines of opposite polarity in the magnetic
reconnection model. The generated shocks should not be associated with 
each other in both models. 

Recently, several papers \citep{Marscher14,Nalewajko11} suggest that particle 
acceleration takes place in the jet flow of a blazar. Emissions from the accelerated particles 
are Doppler boosted by the speeds of the shock and jet flow. In a shock
region where a particle acceleration arises, its speed and angle between the 
moving direction and our line of sight in the jet rest frame are determined with 
statistically random variations. 
Seen in bottom panel of figure~\ref{fig:duration}, there are two types of 
variations. One is the systematic variation caused by the jet flow, the speed of which 
is almost constant. The other is the statistical variation which is caused by the random 
speed and the ejected angle of individual shocks.
If the jet flow changes geometrically, the observed systematic behavior of the 
shot durations can be explained either due to the change in $\delta$ (caused by 
the change of the angle between the jet flow and our line of sight by the 
precession of the jet) or due to the change of the speed of the shock-generated active 
region. The averaged shot durations varied by almost a factor of ten. This
difference is not a result of the statistical variations of individual shots, because
the statistical randomness is diluted by the average of 10 durations. To 
evaluate this fraction of the variation, we estimate the difference of the viewing 
angle $\theta$ assuming that the bulk Lorentz factor $\Gamma$ is constant 
with time. The angle $\theta$ changes from 0$^{\circ}$ to 29$^{\circ}$ for the 
variation of the averaged durations with a factor of ten, assuming
$\Gamma$ of 6.1. This difference of the viewing angle is almost three times 
larger than that of BL~Lac objects \citep{Ajello14}. If the difference of the 
viewing angle is less than 10$^{\circ}$, the $\Gamma$ should be larger than 18.

\subsection{Comparison with Cygnus X-1} \label{sec:cyg}

The existence of flare-like variations has long been known in 
stellar-mass black holes, such as Cygnus~X-1 in the low/hard state, since 
the pioneering work by \citet{Oda71}.
\citet{Negoro94} firstly applied the superposed shot technique to Cygnus~X-1 during its low/hard
state, finding that (1) the average shot at soft X-ray bands
has a rather time-symmetric profile, (2) the profiles are well described
by a sum of exponential functions with time constants of $\sim$0.1~s and
$\sim$1~s both in the rise and decay phases, and (3) the shots in
the rise phase possess a soft energy spectrum, and rapidly harden as the
intensity peaks, resulting in hard X-ray time lags (\cite{Miyamoto93},
and see also \cite{Negoro01}). Since there are apparent similarities
between the shot properties of W2R~1926$+$42 and those of Cygnus~X-1, it may
be interesting to compare them, although the main radiation
mechanisms may differ. (In the case of Cygnus~X-1, there is inverse
Compton scattering radiation with seed photons from an optically thick
disk; see, e.g., \cite{Makishima00}).

We first compare the timescales of shots by scaling these black hole
masses. The spectrum of W2R~1926$+$42 in the optical and infrared bands
contaminates the radiation of its host galaxy. 
The luminosity of the host galaxy of W2R~1926$+$42 in the 
$K_{\rm S}$ band is approximately
1.2$\times10^{44}\;{\rm erg~s}^{-1}\;=\;10^{10.5}\;L_{\odot}$ 
\citep{Edelson13}. The black hole mass of the blazar can then be estimated as 
$10^{7.8}\;M_{\odot}$ by applying the black hole mass-bulge mass
relation \citep{Marconi03}. In this estimation, we assume that the
luminosity from the bulge dominates the observed luminosity of
host galaxy. In comparison, the black hole mass of Cygnus~X-1 is
$\sim15~M_{\odot}$ (see \citep{Orosz11}). The duration of rapid variations of
W2R~1926$+$42 is $\sim 2\;{\tau}_{\rm br}=$7.8${\times}$10$^3$~s, as estimated 
from the break frequency of its PSD. 
This timescale can be scaled to 0.01~s if we assume a mass ratio of 
$10^{-6.6}$ and a Doppler factor of ${\delta}=5.7$ (see \S\ref{sec:mag}).
This scaled timescale is ten times shorter than that of shots seen in 
the black hole X-ray binary Cygnus~X-1. The mean profiles of the hard X-ray
shots are, however, similar, including the asymmetries.

We finally note that the presence of 
two timescales in the shots of Cygnus~X-1 indicates the involvement of 
(at least) two processes: one related to rapid heating and another
to motion of accreting material \citep{Manmoto96,Negoro01}. In fact, the
timescale of $\sim$1~s is too long to explain by a local phenomenon near
the black hole. If the similarity holds, long-term variations of the
shots of W2R~1926$+$42 may partly be related to time variations in
the underlying gas accretion flow that causes the launch of the jets.

We next examine the spectral variations. Asymmetry in the mean profile
of the shots of Cygnus~X-1 is more noticeable in the hard-X-ray band
ranging from 100 to 200~keV \citep{Yamada13}, which is apparently
similar to the rapid variations observed in W2R~1926$+$42. Additionally,
we note again that the shot profile of Cygnus~X-1 contains soft rise and hard decay
features. The shock acceleration scenario may possibly explain some features
of the observed flux and spectral variations of the shots. If accelerated
particles with a given maximum energy emit hard X-ray photons up to
$\sim$200~keV, these particles could be produced rapidly in the
shock, resulting in the rapid spectral hardening near the peak
intensity. A numerical simulation by \citet{Machida03} showed that
particle acceleration near the time of peak flux in Cygnus~X-1 was produced by
magnetic reconnection. \citet{Kirk98} also produced a similar
soft-rise and hard-decay behavior at the frequency of the maximum in the
synchrotron SED in a numerical simulation of shock
acceleration. In stellar black hole binaries, it is also known that
optical lags X-rays variations (e.g., \cite{Spruit02,Gandhi10}). Thus,
spectral changes when shots appear at various wavelengths are very
important to compare observable features of these shots and investigate
their origin.

\section{Conclusion} \label{sec:Conclusion}

We have obtained a continuous optical light curve of the blazar W2R~1926$+$42
with 1-min time resolution with the {\it Kepler} spacecraft. The object exhibits
violent variability and many rapid variations with timescales of hours. 

The power spectrum density (PSD) calculated from the observed light curve 
 cannot be represented a simple power-law function, but instead by a function that 
 combines a power law with a squared Lorentzian function. The best-fit function 
 indicates that the PSD has a break frequency of 
 4.1$^{+0.6}_{-0.5}{\times}10^{-5}$~Hz, which corresponds to 
 0.045${\pm}$0.006~d.

We have detected 195 rapid variations that we describe as shots. 
The amplitude after subtracting long-term baseline components are four times 
larger than the noise level. 
Selected shots show a large diversity in its profiles. An averaged profile can
 be, however, assumed to converge with a universal one reflecting general 
 features of shots, since the observed PSD has a curvature with the break 
 frequency.
 
According to our shot analysis, the 
mean profile produced from detected shots shows several features;
\begin{itemize}
 \item There are three components ranging $-$0.10---0.10~d (component~1),
       $-$0.50---$-$0.15~d (component~2), and 0.10---0.45~d
       (component~3) in the mean profile of the 
       shots. Amplitudes of these components are larger than the
       systematic uncertainty estimated from 
       six mean profiles calculated from different epochs of the light curve.
 \item Component~1 possesses an asymmetric profile, with a faster rise than
       decay and with spiky but smoothly connected behavior near the peak.
       This asymmetry should be caused by the AGN jet physics,
       not the result of stochastic process, because the mean profile of local 
       peaks at simulated noise variation calculated by the Monte-Carlo method 
       can not explain the asymmetry.
 \item Representation by an exponential rise and decay function is
       better than that of another, often used function. E-folding times of
       the rise and decay are 0.043${\pm}$0.001~d and 
       0.061$\pm$0.002~d, respectively.
 \item The timescale estimated from the break frequency shown in the observed PSD
       is consistent with the average 
       of the rise and decay e-folding times of component~1 at the 
       1-sigma confidence level.
 \item The decay phase of the observed mean profile of shots can
       be represented by a simulated light curve based on the
       shock-in-jet scenario. In contrast, the rise phase of the observed
       mean profile is fit by an exponential function rather than by
       alternative functions inferred from past studies with numerical simulations.
\end{itemize}

Durations of the detected shots show a systematic variation with almost a
factor of ten during the monitoring. This systematic variation indicates that
the shots arise associated with each other. 

The shot analysis is also feasible for studying the spectral nature of the
variations, because of the large signal-to-noise ratio. Unfortunately, 
{\it Kepler} performed only one-band monitoring. Spectral and other additional
observational studies are needed to completely understand the mechanism
of rapid variations. 
Additionally, stochastic time-domain analyses, for example 
ARMA-type models, can be applied to the unprecedented high-quality and 
uniformly sampled data provided by {\it Kepler}.

\begin{ack}
The authors thank A.\ Marscher for valuable comments and language editing,
and thank the referee for valuable comments.
This work was supported by Grant-in-Aid for JSPS Fellow (No. 24.1447).
MS is supported by the JSPS Postdoctral Fellowship for Research Abroad.
\end{ack}


\setcounter{table}{1}
\begin{table*}
\centering
\caption{\bf Amplitude and E-folding Timescales of Shots} 
\label{table:ShotElect}
\begin{tabular}{cccccc} \hline \hline
No & Date & Amplitude & Duration & Rise Time & Decay Time \\ \hline
1 & 8.0924 & 152$\pm$17 & 0.042$\pm$0.006 & 0.018$\pm$0.005 & 0.024$\pm$0.001 \\ 
2 & 8.4207 & 82$\pm$17 & 0.096$\pm$0.035 & 0.057$\pm$0.017 & 0.039$\pm$0.018 \\ 
3 & 8.8232 & 71$\pm$14 & 0.090$\pm$0.035 & 0.042$\pm$0.016 & 0.048$\pm$0.020 \\ 
4 & 9.6917 & 69$\pm$14 & 0.068$\pm$0.034 & 0.039$\pm$0.019 & 0.029$\pm$0.015 \\ 
5 & 9.9423 & 110$\pm$14 & 0.129$\pm$0.031 & 0.108$\pm$0.025 & 0.020$\pm$0.005 \\ 
6 & 10.5070 & 96$\pm$20 & 0.110$\pm$0.039 & 0.040$\pm$0.019 & 0.069$\pm$0.020 \\ 
7 & 11.0410 & 90$\pm$12 & 0.066$\pm$0.012 & 0.043$\pm$0.009 & 0.022$\pm$0.003 \\ 
8 & 11.0976 & 92$\pm$12 & 0.060$\pm$0.010 & 0.016$\pm$0.004 & 0.043$\pm$0.005 \\ 
9 & 12.6751 & 96$\pm$12 & 0.125$\pm$0.029 & 0.087$\pm$0.020 & 0.038$\pm$0.009 \\ 
10 & 13.7452 & 78$\pm$13 & 0.052$\pm$0.016 & 0.008$\pm$0.002 & 0.044$\pm$0.014 \\ 
11 & 13.8821 & 79$\pm$13 & 0.078$\pm$0.022 & 0.034$\pm$0.008 & 0.044$\pm$0.014 \\ 
12 & 14.2036 & 118$\pm$13 & 0.182$\pm$0.025 & 0.077$\pm$0.011 & 0.105$\pm$0.014 \\ 
13 & 14.5496 & 123$\pm$12 & 0.109$\pm$0.013 & 0.073$\pm$0.012 & 0.036$\pm$0.001 \\ 
14 & 14.6940 & 120$\pm$12 & 0.160$\pm$0.012 & 0.082$\pm$0.004 & 0.078$\pm$0.008 \\ 
15 & 15.2260 & 115$\pm$13 & 0.143$\pm$0.018 & 0.109$\pm$0.010 & 0.034$\pm$0.007 \\ 
16 & 16.5085 & 69$\pm$14 & 0.120$\pm$0.040 & 0.054$\pm$0.020 & 0.066$\pm$0.020 \\ 
17 & 17.1182 & 98$\pm$14 & 0.171$\pm$0.042 & 0.085$\pm$0.022 & 0.086$\pm$0.019 \\ 
18 & 17.6801 & 245$\pm$25 & 0.249$\pm$0.018 & 0.149$\pm$0.006 & 0.100$\pm$0.012 \\ 
19 & 17.9444 & 179$\pm$25 & 0.225$\pm$0.019 & 0.111$\pm$0.006 & 0.114$\pm$0.013 \\ 
20 & 18.5908 & 145$\pm$17 & 0.143$\pm$0.017 & 0.060$\pm$0.005 & 0.083$\pm$0.013 \\ 
21 & 19.2072 & 136$\pm$16 & 0.121$\pm$0.009 & 0.058$\pm$0.003 & 0.063$\pm$0.007 \\ 
22 & 19.9265 & 101$\pm$16 & 0.093$\pm$0.028 & 0.043$\pm$0.008 & 0.050$\pm$0.020 \\ 
23 & 20.3325 & 74$\pm$13 & 0.064$\pm$0.017 & 0.042$\pm$0.008 & 0.021$\pm$0.009 \\ 
24 & 21.0640 & 118$\pm$13 & 0.123$\pm$0.040 & 0.046$\pm$0.005 & 0.077$\pm$0.035 \\ 
25 & 21.8950 & 94$\pm$14 & 0.066$\pm$0.010 & 0.046$\pm$0.008 & 0.020$\pm$0.002 \\ 
26 & 21.9277 & 93$\pm$14 & 0.034$\pm$0.014 & 0.003$\pm$0.002 & 0.032$\pm$0.012 \\ 
27 & 22.0333 & 216$\pm$13 & 0.098$\pm$0.004 & 0.049$\pm$0.001 & 0.049$\pm$0.003 \\ 
28 & 22.0905 & 229$\pm$13 & 0.138$\pm$0.006 & 0.042$\pm$0.002 & 0.096$\pm$0.003 \\ 
29 & 29.1416 & 80$\pm$13 & 0.123$\pm$0.022 & 0.068$\pm$0.012 & 0.054$\pm$0.010 \\ 
30 & 29.4686 & 181$\pm$17 & 0.293$\pm$0.025 & 0.203$\pm$0.016 & 0.090$\pm$0.009 \\ 
\hline
\end{tabular}
\begin{center}
\footnotesize{Column 1 - Number, 2 - Peak MJD, 3 - Amplitude (c s$^{-1}$), 4 -
 Duration (d), 5 - Rise Time (d), 6 - Decay Time (d)}
\end{center}
\end{table*}

\setcounter{table}{1}
\begin{table*}
\centering
\caption{\bf Continue} 
\label{table:ShotElect}
\begin{tabular}{cccccc} \hline \hline
No & Date & Amplitude & Duration & Rise Time & Decay Time \\ \hline
31 & 29.7492 & 197$\pm$16 & 0.285$\pm$0.023 & 0.122$\pm$0.010 & 0.164$\pm$0.013 \\ 
32 & 30.6490 & 77$\pm$24 & 0.228$\pm$0.118 & 0.169$\pm$0.076 & 0.059$\pm$0.042 \\ 
33 & 31.0815 & 111$\pm$14 & 0.170$\pm$0.032 & 0.086$\pm$0.015 & 0.084$\pm$0.017 \\ 
34 & 31.5045 & 124$\pm$23 & 0.122$\pm$0.052 & 0.068$\pm$0.021 & 0.053$\pm$0.031 \\ 
35 & 32.0017 & 69$\pm$14 & 0.039$\pm$0.017 & 0.020$\pm$0.011 & 0.020$\pm$0.006 \\ 
36 & 32.6583 & 101$\pm$26 & 0.077$\pm$0.040 & 0.031$\pm$0.016 & 0.046$\pm$0.024 \\ 
37 & 32.8293 & 91$\pm$13 & 0.048$\pm$0.013 & 0.034$\pm$0.008 & 0.014$\pm$0.006 \\ 
38 & 32.9961 & 175$\pm$19 & 0.173$\pm$0.022 & 0.070$\pm$0.011 & 0.103$\pm$0.011 \\ 
39 & 33.0929 & 141$\pm$19 & 0.078$\pm$0.011 & 0.035$\pm$0.005 & 0.043$\pm$0.006 \\ 
40 & 33.2706 & 127$\pm$12 & 0.092$\pm$0.013 & 0.034$\pm$0.004 & 0.058$\pm$0.009 \\ 
41 & 33.5451 & 98$\pm$16 & 0.034$\pm$0.007 & 0.024$\pm$0.003 & 0.010$\pm$0.004 \\ 
42 & 33.6371 & 92$\pm$19 & 0.033$\pm$0.011 & 0.020$\pm$0.008 & 0.013$\pm$0.003 \\ 
43 & 33.6623 & 115$\pm$19 & 0.021$\pm$0.004 & 0.012$\pm$0.002 & 0.010$\pm$0.002 \\ 
44 & 33.6861 & 114$\pm$18 & 0.047$\pm$0.008 & 0.027$\pm$0.004 & 0.019$\pm$0.003 \\ 
45 & 33.7175 & 126$\pm$18 & 0.031$\pm$0.007 & 0.019$\pm$0.006 & 0.012$\pm$0.000 \\ 
46 & 33.7604 & 131$\pm$18 & 0.039$\pm$0.004 & 0.027$\pm$0.001 & 0.012$\pm$0.003 \\ 
47 & 33.8646 & 94$\pm$13 & 0.047$\pm$0.010 & 0.017$\pm$0.006 & 0.030$\pm$0.004 \\ 
48 & 33.9300 & 100$\pm$13 & 0.042$\pm$0.009 & 0.012$\pm$0.002 & 0.030$\pm$0.007 \\ 
49 & 33.9926 & 74$\pm$13 & 0.027$\pm$0.005 & 0.019$\pm$0.003 & 0.008$\pm$0.002 \\ 
50 & 34.3714 & 71$\pm$14 & 0.042$\pm$0.010 & 0.015$\pm$0.006 & 0.027$\pm$0.003 \\ 
51 & 34.6043 & 158$\pm$17 & 0.128$\pm$0.006 & 0.077$\pm$0.005 & 0.051$\pm$0.011 \\ 
52 & 34.7221 & 97$\pm$13 & 0.044$\pm$0.010 & 0.038$\pm$0.009 & 0.006$\pm$0.002 \\ 
53 & 34.7977 & 85$\pm$14 & 0.027$\pm$0.004 & 0.013$\pm$0.003 & 0.014$\pm$0.001 \\ 
54 & 34.8495 & 136$\pm$14 & 0.048$\pm$0.001 & 0.026$\pm$0.001 & 0.022$\pm$0.002 \\ 
55 & 34.9490 & 119$\pm$15 & 0.122$\pm$0.020 & 0.052$\pm$0.008 & 0.070$\pm$0.012 \\ 
56 & 35.1363 & 104$\pm$13 & 0.074$\pm$0.004 & 0.036$\pm$0.004 & 0.037$\pm$0.000 \\ 
57 & 35.2146 & 103$\pm$13 & 0.063$\pm$0.009 & 0.030$\pm$0.002 & 0.033$\pm$0.006 \\ 
58 & 35.4176 & 94$\pm$14 & 0.042$\pm$0.010 & 0.031$\pm$0.008 & 0.011$\pm$0.002 \\ 
59 & 35.5763 & 118$\pm$13 & 0.038$\pm$0.013 & 0.014$\pm$0.004 & 0.024$\pm$0.009 \\ 
60 & 35.8828 & 290$\pm$34 & 0.162$\pm$0.028 & 0.059$\pm$0.018 & 0.102$\pm$0.010 \\ 
\hline
\end{tabular}
\end{table*}

\setcounter{table}{1}
\begin{table*}
\centering
\caption{\bf Continue} 
\label{table:ShotElect}
\begin{tabular}{cccccc} \hline \hline
No & Date & Amplitude & Duration & Rise Time & Decay Time \\ \hline
61 & 36.2247 & 307$\pm$26 & 0.147$\pm$0.014 & 0.104$\pm$0.010 & 0.044$\pm$0.003 \\ 
62 & 36.4168 & 236$\pm$26 & 0.094$\pm$0.010 & 0.060$\pm$0.006 & 0.034$\pm$0.004 \\ 
63 & 36.4835 & 248$\pm$26 & 0.186$\pm$0.016 & 0.041$\pm$0.004 & 0.145$\pm$0.012 \\ 
64 & 36.7104 & 179$\pm$26 & 0.087$\pm$0.010 & 0.040$\pm$0.003 & 0.047$\pm$0.007 \\ 
65 & 36.8282 & 141$\pm$26 & 0.080$\pm$0.019 & 0.051$\pm$0.012 & 0.029$\pm$0.007 \\ 
66 & 37.2457 & 122$\pm$23 & 0.058$\pm$0.010 & 0.047$\pm$0.004 & 0.011$\pm$0.006 \\ 
67 & 37.3888 & 139$\pm$25 & 0.074$\pm$0.019 & 0.028$\pm$0.008 & 0.046$\pm$0.010 \\ 
68 & 37.5550 & 101$\pm$21 & 0.034$\pm$0.010 & 0.009$\pm$0.006 & 0.025$\pm$0.004 \\ 
69 & 37.7532 & 110$\pm$13 & 0.044$\pm$0.007 & 0.035$\pm$0.007 & 0.009$\pm$0.000 \\ 
70 & 37.7859 & 96$\pm$13 & 0.034$\pm$0.005 & 0.012$\pm$0.001 & 0.022$\pm$0.004 \\ 
71 & 39.2837 & 116$\pm$12 & 0.040$\pm$0.006 & 0.015$\pm$0.002 & 0.024$\pm$0.003 \\ 
72 & 39.4315 & 156$\pm$15 & 0.059$\pm$0.004 & 0.030$\pm$0.001 & 0.030$\pm$0.003 \\ 
73 & 39.4771 & 186$\pm$15 & 0.058$\pm$0.004 & 0.036$\pm$0.003 & 0.022$\pm$0.001 \\ 
74 & 39.5990 & 111$\pm$18 & 0.049$\pm$0.008 & 0.035$\pm$0.005 & 0.014$\pm$0.002 \\ 
75 & 39.6883 & 73$\pm$19 & 0.020$\pm$0.007 & 0.004$\pm$0.002 & 0.015$\pm$0.005 \\ 
76 & 40.1623 & 119$\pm$18 & 0.114$\pm$0.020 & 0.051$\pm$0.006 & 0.063$\pm$0.015 \\ 
77 & 40.4975 & 108$\pm$15 & 0.135$\pm$0.028 & 0.051$\pm$0.011 & 0.084$\pm$0.017 \\ 
78 & 40.7583 & 70$\pm$14 & 0.018$\pm$0.005 & 0.009$\pm$0.003 & 0.009$\pm$0.002 \\ 
79 & 40.8040 & 72$\pm$15 & 0.024$\pm$0.002 & 0.018$\pm$0.001 & 0.007$\pm$0.001 \\ 
80 & 40.8387 & 154$\pm$18 & 0.040$\pm$0.007 & 0.016$\pm$0.003 & 0.024$\pm$0.004 \\ 
81 & 41.5171 & 159$\pm$13 & 0.051$\pm$0.005 & 0.022$\pm$0.002 & 0.030$\pm$0.002 \\ 
82 & 41.5716 & 91$\pm$13 & 0.023$\pm$0.003 & 0.018$\pm$0.002 & 0.005$\pm$0.002 \\ 
83 & 41.6608 & 172$\pm$35 & 0.083$\pm$0.020 & 0.044$\pm$0.012 & 0.039$\pm$0.008 \\ 
84 & 41.7882 & 213$\pm$14 & 0.054$\pm$0.004 & 0.024$\pm$0.002 & 0.029$\pm$0.002 \\ 
85 & 41.8590 & 256$\pm$14 & 0.061$\pm$0.009 & 0.040$\pm$0.007 & 0.021$\pm$0.002 \\ 
86 & 42.0613 & 161$\pm$13 & 0.044$\pm$0.004 & 0.023$\pm$0.004 & 0.021$\pm$0.000 \\ 
87 & 42.1219 & 177$\pm$13 & 0.048$\pm$0.000 & 0.032$\pm$0.001 & 0.016$\pm$0.001 \\ 
88 & 42.4720 & 114$\pm$14 & 0.039$\pm$0.002 & 0.011$\pm$0.001 & 0.028$\pm$0.001 \\ 
89 & 42.5769 & 130$\pm$15 & 0.044$\pm$0.009 & 0.028$\pm$0.004 & 0.016$\pm$0.005 \\ 
90 & 42.6941 & 124$\pm$16 & 0.034$\pm$0.003 & 0.014$\pm$0.001 & 0.020$\pm$0.002 \\ 
\hline
\end{tabular}
\end{table*}

\setcounter{table}{1}
\begin{table*}
\centering
\caption{\bf Continue} 
\label{table:ShotElect}
\begin{tabular}{cccccc} \hline \hline
No & Date & Amplitude & Duration & Rise Time & Decay Time \\ \hline
91 & 42.7636 & 388$\pm$33 & 0.048$\pm$0.004 & 0.029$\pm$0.002 & 0.019$\pm$0.002 \\ 
92 & 42.7813 & 455$\pm$33 & 0.069$\pm$0.009 & 0.017$\pm$0.001 & 0.052$\pm$0.008 \\ 
93 & 42.9631 & 192$\pm$23 & 0.118$\pm$0.014 & 0.038$\pm$0.004 & 0.081$\pm$0.010 \\ 
94 & 43.1218 & 76$\pm$17 & 0.063$\pm$0.012 & 0.018$\pm$0.006 & 0.045$\pm$0.007 \\ 
95 & 43.2397 & 126$\pm$17 & 0.075$\pm$0.011 & 0.026$\pm$0.005 & 0.049$\pm$0.006 \\ 
96 & 43.5796 & 137$\pm$13 & 0.093$\pm$0.013 & 0.058$\pm$0.008 & 0.035$\pm$0.005 \\ 
97 & 44.2791 & 125$\pm$12 & 0.053$\pm$0.006 & 0.036$\pm$0.006 & 0.017$\pm$0.000 \\ 
98 & 44.3458 & 95$\pm$12 & 0.043$\pm$0.006 & 0.031$\pm$0.004 & 0.011$\pm$0.003 \\ 
99 & 44.5379 & 103$\pm$25 & 0.044$\pm$0.011 & 0.024$\pm$0.007 & 0.020$\pm$0.004 \\ 
100 & 44.6925 & 91$\pm$13 & 0.034$\pm$0.007 & 0.024$\pm$0.004 & 0.010$\pm$0.002 \\ 
101 & 44.8451 & 160$\pm$16 & 0.067$\pm$0.008 & 0.034$\pm$0.005 & 0.033$\pm$0.003 \\ 
102 & 45.0283 & 213$\pm$19 & 0.101$\pm$0.016 & 0.070$\pm$0.009 & 0.031$\pm$0.007 \\ 
103 & 45.1584 & 113$\pm$18 & 0.043$\pm$0.018 & 0.020$\pm$0.013 & 0.022$\pm$0.004 \\ 
104 & 45.3839 & 237$\pm$23 & 0.097$\pm$0.008 & 0.045$\pm$0.003 & 0.052$\pm$0.005 \\ 
105 & 45.4656 & 380$\pm$23 & 0.109$\pm$0.002 & 0.043$\pm$0.001 & 0.066$\pm$0.001 \\ 
106 & 45.8940 & 189$\pm$19 & 0.079$\pm$0.006 & 0.049$\pm$0.003 & 0.030$\pm$0.003 \\ 
107 & 45.9533 & 158$\pm$19 & 0.056$\pm$0.007 & 0.031$\pm$0.004 & 0.025$\pm$0.003 \\ 
108 & 46.1154 & 190$\pm$18 & 0.037$\pm$0.005 & 0.025$\pm$0.004 & 0.012$\pm$0.001 \\ 
109 & 46.2414 & 221$\pm$15 & 0.053$\pm$0.001 & 0.014$\pm$0.000 & 0.039$\pm$0.000 \\ 
110 & 46.4771 & 131$\pm$21 & 0.052$\pm$0.006 & 0.018$\pm$0.001 & 0.034$\pm$0.005 \\ 
111 & 46.9239 & 186$\pm$22 & 0.183$\pm$0.026 & 0.070$\pm$0.008 & 0.113$\pm$0.017 \\ 
112 & 47.2290 & 171$\pm$18 & 0.029$\pm$0.006 & 0.014$\pm$0.004 & 0.016$\pm$0.002 \\ 
113 & 47.3346 & 118$\pm$15 & 0.050$\pm$0.003 & 0.010$\pm$0.002 & 0.039$\pm$0.001 \\ 
114 & 47.4749 & 80$\pm$16 & 0.025$\pm$0.008 & 0.014$\pm$0.006 & 0.011$\pm$0.002 \\ 
115 & 47.6330 & 129$\pm$21 & 0.032$\pm$0.007 & 0.013$\pm$0.003 & 0.019$\pm$0.003 \\ 
116 & 47.7160 & 192$\pm$23 & 0.038$\pm$0.006 & 0.024$\pm$0.004 & 0.014$\pm$0.002 \\ 
117 & 47.7419 & 175$\pm$23 & 0.072$\pm$0.010 & 0.035$\pm$0.005 & 0.037$\pm$0.005 \\ 
118 & 47.7835 & 147$\pm$23 & 0.040$\pm$0.009 & 0.022$\pm$0.003 & 0.018$\pm$0.006 \\ 
119 & 47.9074 & 210$\pm$18 & 0.053$\pm$0.004 & 0.024$\pm$0.002 & 0.028$\pm$0.002 \\ 
120 & 47.9749 & 140$\pm$18 & 0.032$\pm$0.005 & 0.016$\pm$0.002 & 0.016$\pm$0.002 \\ 
\hline
\end{tabular}
\end{table*}

\setcounter{table}{1}
\begin{table*}
\centering
\caption{\bf Continue} 
\label{table:ShotElect}
\begin{tabular}{cccccc} \hline \hline
No & Date & Amplitude & Duration & Rise Time & Decay Time \\ \hline
121 & 48.0355 & 149$\pm$18 & 0.049$\pm$0.006 & 0.030$\pm$0.003 & 0.019$\pm$0.003 \\ 
122 & 48.2214 & 82$\pm$17 & 0.059$\pm$0.008 & 0.028$\pm$0.005 & 0.031$\pm$0.003 \\ 
123 & 48.5402 & 85$\pm$13 & 0.062$\pm$0.011 & 0.035$\pm$0.005 & 0.027$\pm$0.006 \\ 
124 & 49.2615 & 137$\pm$16 & 0.146$\pm$0.008 & 0.042$\pm$0.005 & 0.104$\pm$0.004 \\ 
125 & 49.5306 & 162$\pm$17 & 0.102$\pm$0.015 & 0.057$\pm$0.007 & 0.045$\pm$0.008 \\ 
126 & 49.7492 & 109$\pm$14 & 0.130$\pm$0.013 & 0.038$\pm$0.005 & 0.093$\pm$0.008 \\ 
127 & 49.9624 & 119$\pm$15 & 0.159$\pm$0.006 & 0.093$\pm$0.001 & 0.065$\pm$0.005 \\ 
128 & 50.1661 & 100$\pm$15 & 0.146$\pm$0.031 & 0.087$\pm$0.017 & 0.058$\pm$0.014 \\ 
129 & 50.5918 & 88$\pm$16 & 0.117$\pm$0.027 & 0.062$\pm$0.021 & 0.055$\pm$0.006 \\ 
130 & 50.8411 & 100$\pm$13 & 0.157$\pm$0.020 & 0.067$\pm$0.006 & 0.091$\pm$0.014 \\ 
131 & 51.4016 & 120$\pm$16 & 0.109$\pm$0.015 & 0.043$\pm$0.008 & 0.066$\pm$0.008 \\ 
132 & 52.1522 & 125$\pm$15 & 0.105$\pm$0.016 & 0.034$\pm$0.007 & 0.071$\pm$0.009 \\ 
133 & 52.5371 & 134$\pm$18 & 0.121$\pm$0.018 & 0.073$\pm$0.012 & 0.048$\pm$0.006 \\ 
134 & 52.6133 & 123$\pm$18 & 0.112$\pm$0.015 & 0.037$\pm$0.005 & 0.075$\pm$0.009 \\ 
135 & 53.0431 & 129$\pm$16 & 0.090$\pm$0.010 & 0.040$\pm$0.006 & 0.050$\pm$0.004 \\ 
136 & 53.2638 & 215$\pm$16 & 0.158$\pm$0.009 & 0.095$\pm$0.007 & 0.063$\pm$0.002 \\ 
137 & 53.6228 & 235$\pm$28 & 0.089$\pm$0.007 & 0.038$\pm$0.003 & 0.051$\pm$0.004 \\ 
138 & 53.8298 & 118$\pm$18 & 0.081$\pm$0.013 & 0.030$\pm$0.005 & 0.051$\pm$0.008 \\ 
139 & 53.9136 & 104$\pm$18 & 0.078$\pm$0.006 & 0.025$\pm$0.000 & 0.053$\pm$0.006 \\ 
140 & 54.1772 & 156$\pm$14 & 0.099$\pm$0.007 & 0.024$\pm$0.003 & 0.076$\pm$0.004 \\ 
141 & 54.9360 & 170$\pm$16 & 0.170$\pm$0.010 & 0.078$\pm$0.005 & 0.092$\pm$0.004 \\ 
142 & 55.1928 & 135$\pm$14 & 0.115$\pm$0.032 & 0.038$\pm$0.010 & 0.077$\pm$0.022 \\ 
143 & 55.5170 & 122$\pm$18 & 0.106$\pm$0.013 & 0.050$\pm$0.008 & 0.056$\pm$0.005 \\ 
144 & 55.6920 & 205$\pm$17 & 0.165$\pm$0.015 & 0.052$\pm$0.005 & 0.114$\pm$0.010 \\ 
145 & 55.8432 & 221$\pm$17 & 0.205$\pm$0.016 & 0.102$\pm$0.008 & 0.104$\pm$0.008 \\ 
146 & 55.9931 & 222$\pm$17 & 0.076$\pm$0.004 & 0.061$\pm$0.003 & 0.015$\pm$0.001 \\ 
147 & 56.0367 & 227$\pm$17 & 0.130$\pm$0.010 & 0.071$\pm$0.005 & 0.059$\pm$0.004 \\ 
148 & 56.1211 & 216$\pm$17 & 0.161$\pm$0.013 & 0.122$\pm$0.010 & 0.039$\pm$0.003 \\ 
149 & 56.3207 & 150$\pm$17 & 0.079$\pm$0.010 & 0.055$\pm$0.007 & 0.024$\pm$0.003 \\ 
150 & 56.3765 & 164$\pm$17 & 0.199$\pm$0.014 & 0.054$\pm$0.006 & 0.146$\pm$0.009 \\ 
\hline
\end{tabular}
\end{table*}

\setcounter{table}{1}
\begin{table*}
\centering
\caption{\bf Continue} 
\label{table:ShotElect}
\begin{tabular}{cccccc} \hline \hline
No & Date & Amplitude & Duration & Rise Time & Decay Time \\ \hline
151 & 56.7662 & 102$\pm$14 & 0.169$\pm$0.023 & 0.147$\pm$0.013 & 0.023$\pm$0.010 \\ 
152 & 56.9855 & 124$\pm$14 & 0.131$\pm$0.014 & 0.087$\pm$0.009 & 0.045$\pm$0.005 \\ 
153 & 57.5746 & 103$\pm$14 & 0.137$\pm$0.024 & 0.067$\pm$0.010 & 0.070$\pm$0.013 \\ 
154 & 58.0909 & 71$\pm$12 & 0.144$\pm$0.034 & 0.060$\pm$0.013 & 0.084$\pm$0.022 \\ 
155 & 59.0322 & 99$\pm$13 & 0.096$\pm$0.010 & 0.078$\pm$0.009 & 0.018$\pm$0.000 \\ 
156 & 59.0772 & 111$\pm$13 & 0.112$\pm$0.022 & 0.019$\pm$0.004 & 0.093$\pm$0.018 \\ 
157 & 59.2849 & 116$\pm$13 & 0.171$\pm$0.018 & 0.115$\pm$0.017 & 0.056$\pm$0.000 \\ 
158 & 59.4191 & 94$\pm$13 & 0.104$\pm$0.033 & 0.007$\pm$0.016 & 0.097$\pm$0.017 \\ 
159 & 59.8871 & 166$\pm$12 & 0.226$\pm$0.020 & 0.144$\pm$0.014 & 0.082$\pm$0.006 \\ 
160 & 60.0137 & 205$\pm$12 & 0.270$\pm$0.015 & 0.086$\pm$0.005 & 0.184$\pm$0.010 \\ 
161 & 60.2160 & 134$\pm$12 & 0.177$\pm$0.018 & 0.095$\pm$0.008 & 0.083$\pm$0.009 \\ 
162 & 60.5607 & 113$\pm$13 & 0.092$\pm$0.017 & 0.043$\pm$0.007 & 0.048$\pm$0.010 \\ 
163 & 60.8897 & 88$\pm$12 & 0.158$\pm$0.033 & 0.107$\pm$0.026 & 0.051$\pm$0.008 \\ 
164 & 61.1383 & 195$\pm$13 & 0.196$\pm$0.009 & 0.076$\pm$0.003 & 0.119$\pm$0.006 \\ 
165 & 61.5469 & 124$\pm$17 & 0.219$\pm$0.028 & 0.138$\pm$0.008 & 0.081$\pm$0.019 \\ 
166 & 62.6388 & 193$\pm$23 & 0.192$\pm$0.032 & 0.066$\pm$0.011 & 0.125$\pm$0.021 \\ 
167 & 63.0045 & 126$\pm$17 & 0.141$\pm$0.021 & 0.072$\pm$0.016 & 0.069$\pm$0.006 \\ 
168 & 63.1571 & 189$\pm$17 & 0.223$\pm$0.003 & 0.068$\pm$0.005 & 0.154$\pm$0.007 \\ 
169 & 63.3805 & 146$\pm$17 & 0.125$\pm$0.024 & 0.074$\pm$0.014 & 0.051$\pm$0.010 \\ 
170 & 63.4854 & 132$\pm$17 & 0.120$\pm$0.020 & 0.055$\pm$0.007 & 0.066$\pm$0.013 \\ 
171 & 64.4955 & 111$\pm$15 & 0.367$\pm$0.074 & 0.079$\pm$0.020 & 0.288$\pm$0.054 \\ 
172 & 67.5347 & 175$\pm$22 & 0.241$\pm$0.026 & 0.157$\pm$0.011 & 0.084$\pm$0.015 \\ 
173 & 67.9290 & 293$\pm$20 & 0.313$\pm$0.022 & 0.137$\pm$0.010 & 0.176$\pm$0.012 \\ 
174 & 68.1347 & 204$\pm$20 & 0.193$\pm$0.019 & 0.096$\pm$0.009 & 0.097$\pm$0.009 \\ 
175 & 68.6238 & 351$\pm$20 & 0.306$\pm$0.016 & 0.098$\pm$0.009 & 0.208$\pm$0.007 \\ 
176 & 71.0853 & 288$\pm$16 & 0.339$\pm$0.015 & 0.120$\pm$0.006 & 0.219$\pm$0.009 \\ 
177 & 71.9496 & 188$\pm$15 & 0.207$\pm$0.021 & 0.107$\pm$0.013 & 0.100$\pm$0.008 \\ 
178 & 72.1070 & 176$\pm$15 & 0.231$\pm$0.020 & 0.102$\pm$0.009 & 0.129$\pm$0.012 \\ 
179 & 72.3651 & 189$\pm$16 & 0.388$\pm$0.032 & 0.269$\pm$0.021 & 0.119$\pm$0.010 \\ 
180 & 72.5136 & 191$\pm$16 & 0.229$\pm$0.019 & 0.066$\pm$0.005 & 0.163$\pm$0.014 \\ 
\hline
\end{tabular}
\end{table*}

\setcounter{table}{1}
\begin{table*}
\centering
\caption{\bf Continue} 
\label{table:ShotElect}
\begin{tabular}{cccccc} \hline \hline
No & Date & Amplitude & Duration & Rise Time & Decay Time \\ \hline
181 & 72.7595 & 241$\pm$16 & 0.408$\pm$0.027 & 0.181$\pm$0.012 & 0.226$\pm$0.015 \\ 
182 & 73.1014 & 254$\pm$16 & 0.306$\pm$0.019 & 0.158$\pm$0.010 & 0.148$\pm$0.009 \\ 
183 & 73.4229 & 295$\pm$15 & 0.289$\pm$0.016 & 0.238$\pm$0.013 & 0.051$\pm$0.003 \\ 
184 & 73.4856 & 259$\pm$15 & 0.316$\pm$0.020 & 0.094$\pm$0.006 & 0.223$\pm$0.015 \\ 
185 & 74.0488 & 110$\pm$14 & 0.126$\pm$0.015 & 0.062$\pm$0.007 & 0.064$\pm$0.008 \\ 
186 & 74.4514 & 93$\pm$13 & 0.102$\pm$0.017 & 0.030$\pm$0.006 & 0.072$\pm$0.011 \\ 
187 & 74.9990 & 87$\pm$13 & 0.185$\pm$0.039 & 0.080$\pm$0.016 & 0.106$\pm$0.023 \\ 
188 & 76.0683 & 80$\pm$12 & 0.210$\pm$0.059 & 0.151$\pm$0.046 & 0.059$\pm$0.013 \\ 
189 & 76.4395 & 75$\pm$13 & 0.105$\pm$0.031 & 0.063$\pm$0.022 & 0.042$\pm$0.010 \\ 
190 & 77.0226 & 108$\pm$12 & 0.102$\pm$0.017 & 0.060$\pm$0.008 & 0.042$\pm$0.008 \\ 
191 & 77.1561 & 100$\pm$12 & 0.295$\pm$0.057 & 0.105$\pm$0.032 & 0.190$\pm$0.025 \\ 
192 & 80.7578 & 109$\pm$13 & 0.391$\pm$0.112 & 0.232$\pm$0.064 & 0.158$\pm$0.048 \\ 
193 & 85.2674 & 87$\pm$13 & 0.497$\pm$0.137 & 0.230$\pm$0.079 & 0.267$\pm$0.058 \\ 
194 & 86.1263 & 73$\pm$12 & 0.128$\pm$0.018 & 0.085$\pm$0.017 & 0.042$\pm$0.001 \\ 
195 & 95.1884 & 95$\pm$14 & 0.706$\pm$0.317 & 0.275$\pm$0.127 & 0.431$\pm$0.191 \\ 
\hline
\end{tabular}
\end{table*}

\clearpage

\setcounter{figure}{11}
\begin{figure*}
\begin{center}
\begin{tabular}{c}
\includegraphics[width=16cm]{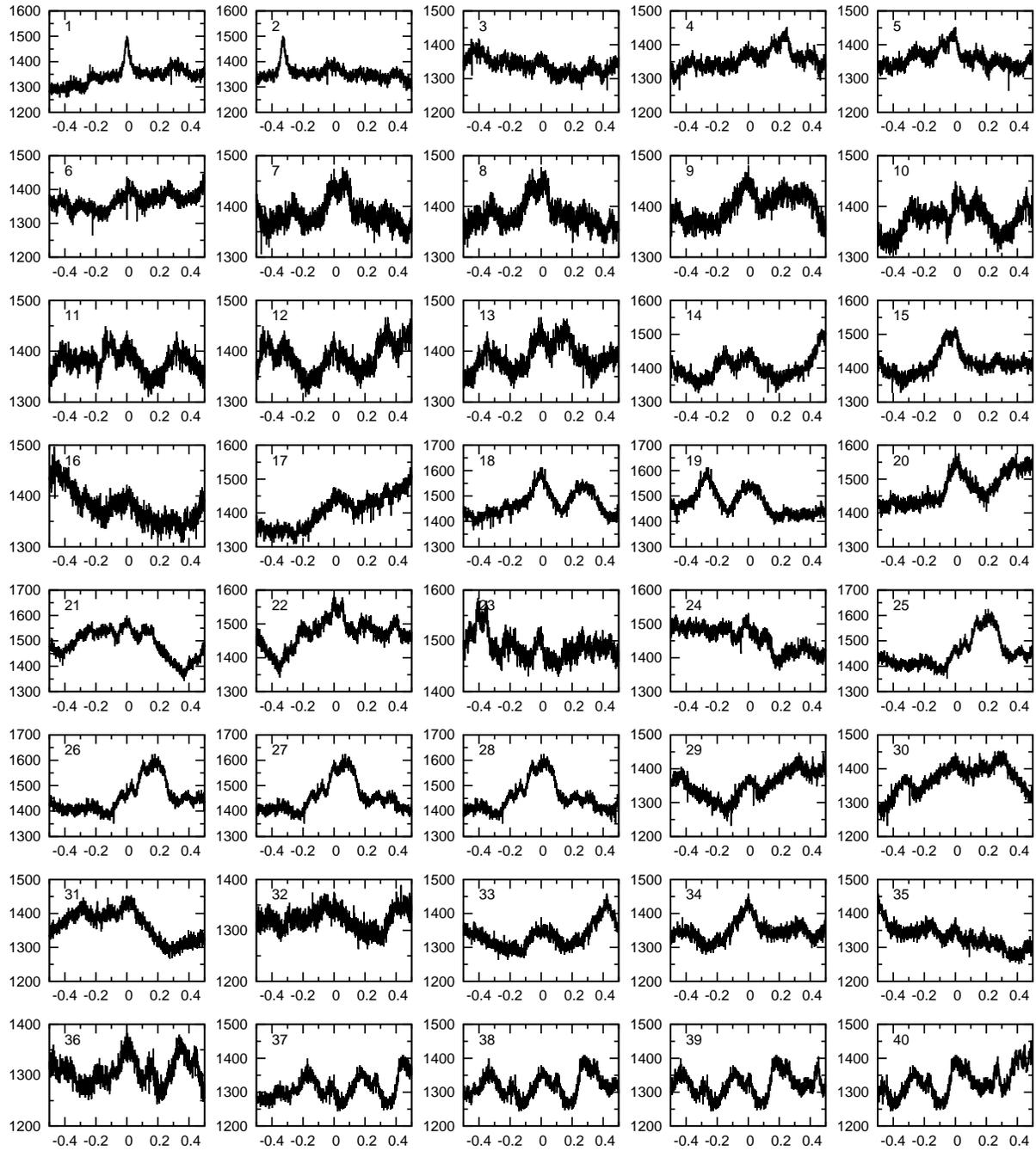}
\end{tabular}
 \caption{Light curve of each detected shot obtained by the {\it Kepler} spacecraft over the entire
 Quarter~14 period. The object was monitored for 100~d with 1-minute time resolution. Each shot
 is numbered at the left top. The number of each shot is listed in table~\ref{table:ShotElect}.}
 \label{fig:lcElect}
\end{center}
\end{figure*}

\setcounter{figure}{11}
\begin{figure*}
\begin{center}
\begin{tabular}{c}
\includegraphics[width=16cm]{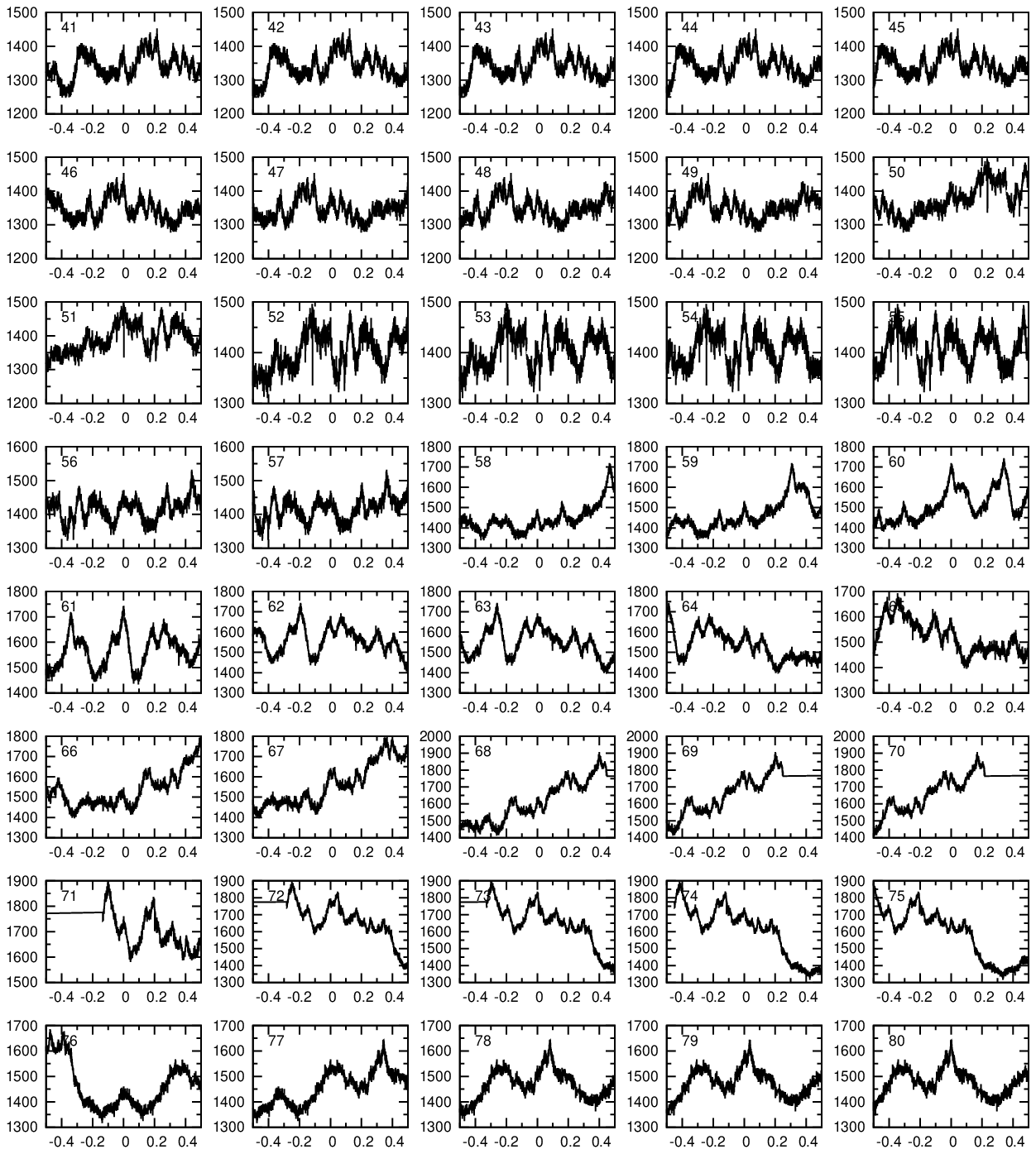}
\end{tabular}
\caption{(Continued)}
\end{center}
\end{figure*}

\setcounter{figure}{11}
\begin{figure*}
\begin{center}
\begin{tabular}{c}
\includegraphics[width=16cm]{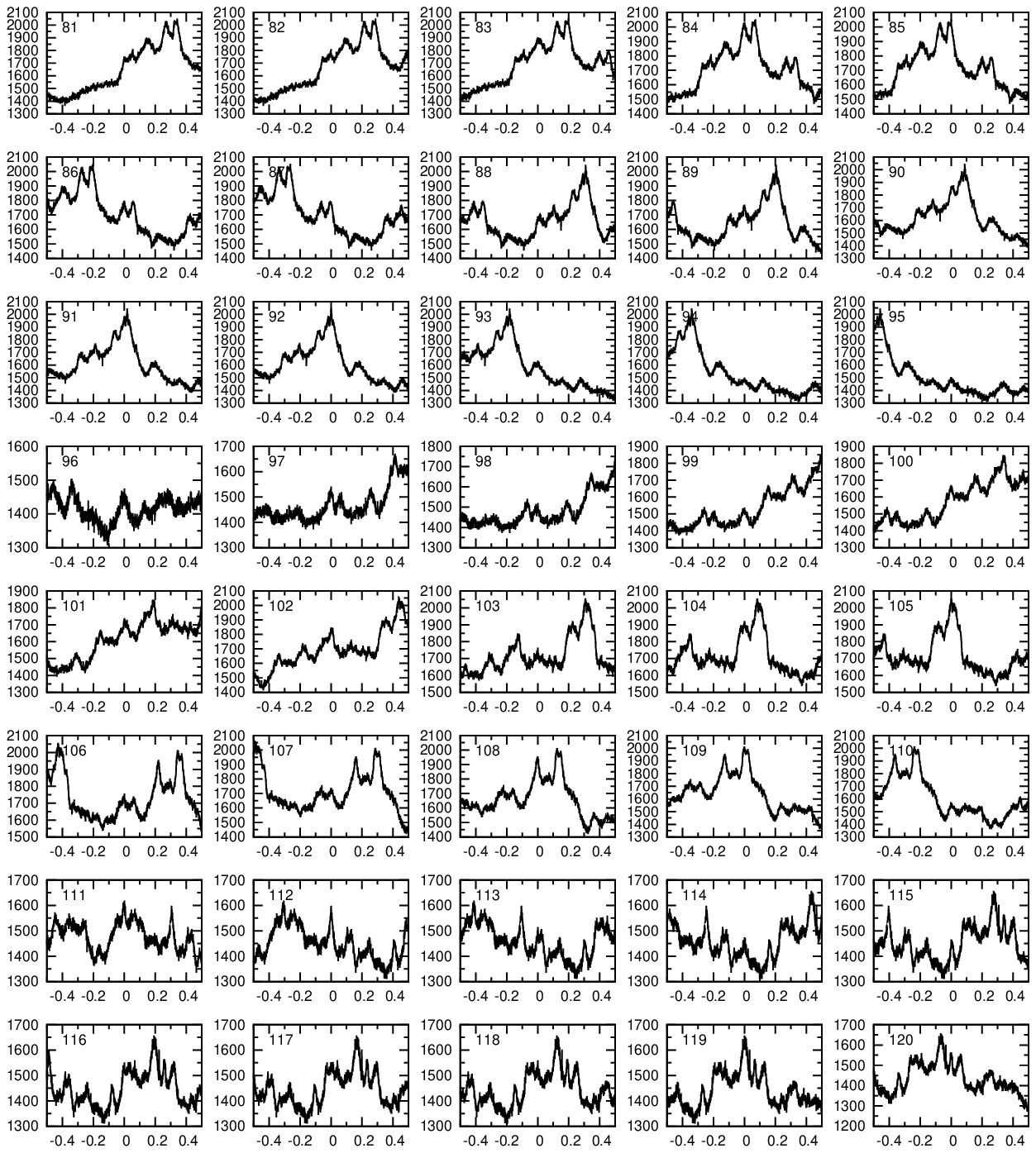}
\end{tabular}
\caption{(Continued)}
\end{center}
\end{figure*}

\setcounter{figure}{11}
\begin{figure*}
\begin{center}
\begin{tabular}{c}
\includegraphics[width=16cm]{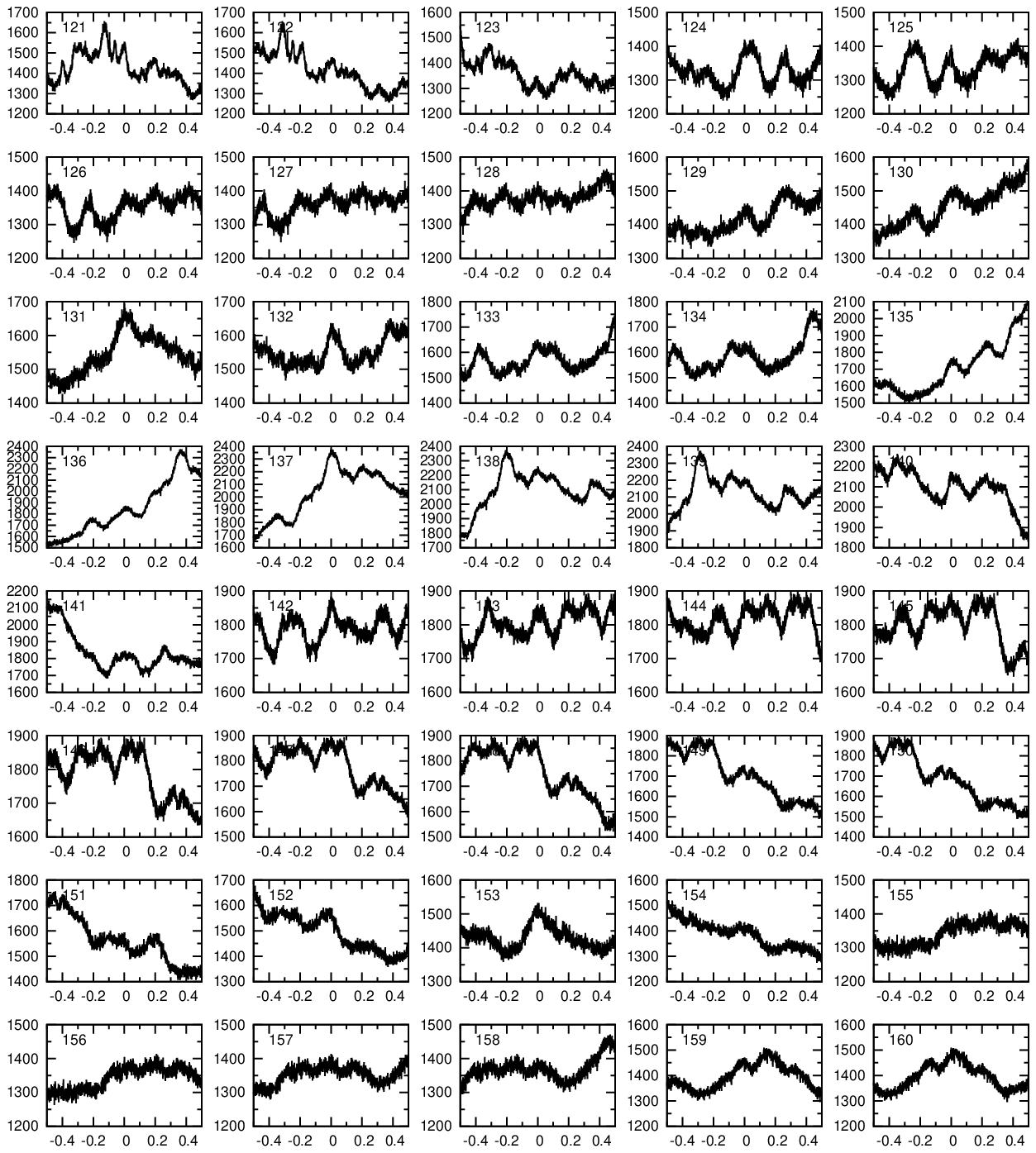}
\end{tabular}
\caption{(Continued)}
\end{center}
\end{figure*}

\setcounter{figure}{11}
\begin{figure*}
\begin{center}
\begin{tabular}{c}
\includegraphics[width=16cm]{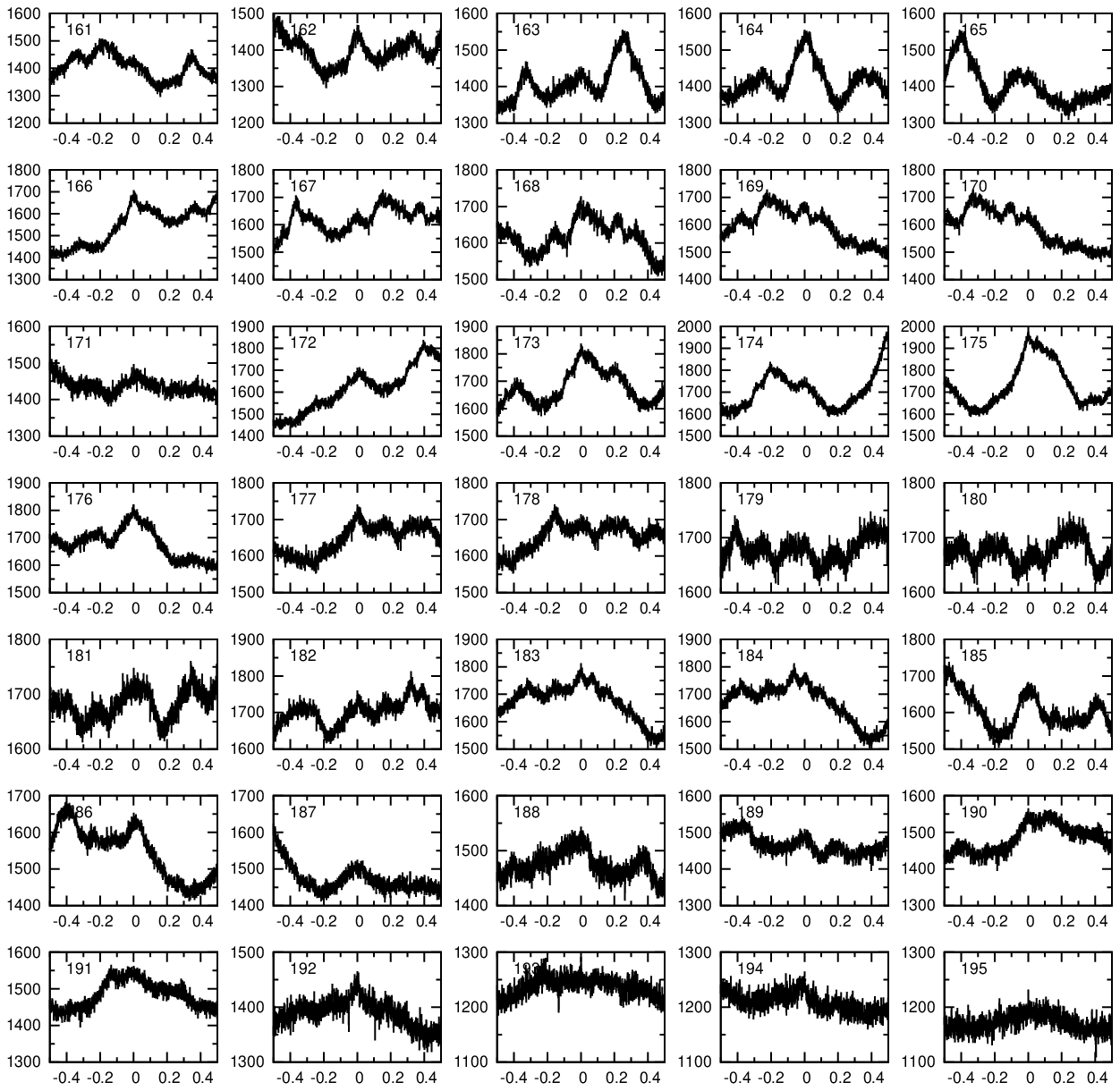}
\end{tabular}
\caption{(Continued)}
\end{center}
\end{figure*}


\begin{thebibliography}{}

\bibitem[Abdo et~al.(2010)]{Abdo10} Abdo, A. A., et~al. 2010,
                {\apj}, 722, 520

\bibitem[Abdo et~al.(2011)]{Abdo11} Abdo, A. A., et~al. 2011,
                {\apjl}, 733, L26

\bibitem[Ackermann et~al.(2016)]{Ackermann16} Ackermann, M., et~al. 2016,
                {\apjl}, 824, L20

\bibitem[Aharonian et~al.(2007)]{Aharonian07} Aharonian, F., et~al. 2007,
                {\apjl}, 664, L71

\bibitem[Ajello et~al.(2014)]{Ajello14} Ajello, M., et~al. 2014,
                {\apj}, 780, 73

\bibitem[Antonucci(1993)]{Antonucci93} Antonucci, R. 1993, {\araa}, 31, 473

\bibitem[Blandford \& K\"onigl(1979)]{Blandford79} Blandford, R. D. \&
                K\"onigl, A. 1979, {\apj}, 232, 34

\bibitem[Borucki et~al.(2010)]{Borucki10} Borucki, W., et~al. 2010,
                Sci, 327, 977

\bibitem[B\"ottcher \& Dermer(2010)]{Bottcher10} B\"ottcher, M., \&
                Dermer, C. D. 2010, {\apj}, 711, 445

\bibitem[Carini et~al.(1990)]{Carini90} Carini, M., Miller, H. R., \&
                Goodrich, B. D. 1990, {\aj}, 100, 347

\bibitem[Chang \& Refsdal(1979)]{Chang79} Chang, K., \& Refsdal, S. 1979,
                {\nat}, 282, 561

\bibitem[Edelson \& Malkan(2012)]{Edelson12} Edelson, R., \& Malkan, M. 2012,
                {\apj}, 751, 52

\bibitem[Edelson et~al.(2013)]{Edelson13} Edelson, R., et~al. 2013,
                {\apj}, 766, 16

\bibitem[Efron \& Tibshirani(1979)]{Efron79} Efron, B. 1979
                Ann. Stat., 7, 1

\bibitem[Efron(1994)]{Efron94} Efron, B., \& Tibshirani, R. 1994, An Introduction to the Bootstrap (New York: Chapman \& Hall CRC Monographs on Statistics \& Applied Probability; Taylor and Francis

\bibitem[Gandhi et~al.(2010)]{Gandhi10} Gandhi, P., et~al. 2010,
                {\mnras}, 407, 2166

\bibitem[Ghisellini et~al.(2010)]{Ghisellini10} Ghisellini, G., et~al. 2010,
                {\mnras}, 402, 497

\bibitem[Giannios et~al.(2009)]{Giannios09} Giannios, D., Uzdensky,
                D. A., \& Begelman, M. C. 2009, {\mnras}, 395, L29

\bibitem[Jenkins et~al.(2010)]{Jenkins10} Jenkins, J., et~al. 2010,
                {\apjl}, 713, L87

\bibitem[Kataoka et~al.(2001)]{Kataoka01} Kataoka, J., et~al. 2001,
                {\apj}, 560, 659

\bibitem[Kirk, Rieger \& Mastichiadis(1998)]{Kirk98} Kirk, J. G.,
                Rieger, F. M., \& Mastichiadis, A. 1998, {\aap}, 333, 452

\bibitem[Koch et~al.(2010)]{Koch10} Koch, D. G., et~al. 2010, {\apj}, 713, L79

\bibitem[Kusunose, Takahara \& Li(2000)]{Kusunose00} Kusunose, M.,
                Takahara, F., \& Li, H. 2000, {\apj}, 536, 299

\bibitem[Machida \& Matsumoto(2003)]{Machida03} Machida, M., \&
                Matsumoto, R. 2003, {\apj}, 585, 429

\bibitem[Makishima et~al.(2000)]{Makishima00} Makishima, K., et~al.
                2000, {\apj}, 535, 632

\bibitem[Manmoto et~al.(1996)]{Manmoto96} Manmoto, T., Takeuchi, M.,
                Mineshige, S., Matsumoto, R., Negoro, H. 1996, {\apjl},
                464, L135

\bibitem[Mann \& Whitney(1947)]{Mann47} Mann, H. B., \& Whitney,
                D. R. 1947, Annals of Mathematical Statistics, 18, 50

\bibitem[Marconi \& Hunt(2003)]{Marconi03} Marconi, A., \& Hunt,
                L. K. 2003, {\apjl}, 589, L21 

\bibitem[Marscher \& Gear(1985)]{Marscher85} Marscher, A. P., \& Gear,
                W. K. 1985, {\apj}, 298, 114

\bibitem[Marscher(2014)]{Marscher14} Marscher, A. P. 2014, {\apj} 780, 87

\bibitem[Miyamoto et~al.(1993)]{Miyamoto93} Miyamoto, S., Iga, S.,
                Kitamoto, S., \& Kamado, Y. 1993, {\apjl}, 403, L39

\bibitem[Mohan et~al.(2016)]{Mohan16} Mohan, P., Gupta, A. C.,
                Bachev, R., \& Strigachev, A. 2016, {\mnras}, 456, 654
                
\bibitem[Nalewajko et~al.(2011)]{Nalewajko11} Nalewajko, K., Niannios, D.,
               Begelman, M. C., Uzdensky, D. A., \& Sikora, M. 2011,
               {\mnras} 413, 333

\bibitem[Nalewajko(2013)]{Nalewajko13} Nalewajko, K. 2013, {\mnras}, 430, 1324

\bibitem[Negoro et~al.(1994)]{Negoro94} Negoro, H., Miyamoto, S., \&
                Kitamoto, S. 1994, {\apjl}, 423, L127

\bibitem[Negoro et~al.(2001)]{Negoro01} Negoro, H., Kitamoto, S., \&
                Mineshige, S. 2001, {\apj}, 554, 528

\bibitem[Negoro et~al.(2002)]{Negoro02} Negoro, H., \&
                Mineshige, S. 2002, {\pasj}, 54, L69

\bibitem[Oda et~al.(1971)]{Oda71} Oda, M., Gorenstein, P., Gursky, H.,
                Kellogg, E., Schreier, E., Tananbaum, H., \& Giacconi,
                R. 1971, {\apjl}, 166, L1

\bibitem[Orosz et~al.(2011)]{Orosz11} Orosz, J. A., McClintock,
                J. E., Aufdenberg, J. P., Remillard, R. A.,
                Reid, M. J., Narayan, R., \& Gou, L.
                2011, {\apj}, 742, 84

\bibitem[Papadakis \& Lawrence(1993)]{Papadakis93} Papadakis, I. E.,
                \& Lawrence, A. 1993, {\mnras}, 261, 612 

\bibitem[Quirrenbach et~al.(1992)]{Quirrenbach92} Quirrenbach, A.,
                et~al. 1992, {\aap}, 258, 279 

\bibitem[Sasada et~al.(2008)]{Sasada08} Sasada, M., et~al. 2008,
                {\pasj}, 60, 37

\bibitem[Sasada et~al.(2010)]{Sasada10} Sasada, M., et~al. 2010,
                {\pasj}, 62, 645

\bibitem[Saito et~al.(2013)]{Saito13} Saito, S., Stawarz, \L., Tanaka,
                Y. T., Takahashi, T., Madejski, G., \& D'Ammando, F. 2013,
                {\apjl}, 766, L11

\bibitem[Spada et~al.(2001)]{Spada01} Spada, M., Ghisellini, G., Lazzati,
                D., \& Celotti, A. 2001, {\mnras}, 325, 1559

\bibitem[Spruit \& Kanbach(2002)]{Spruit02} Spruit, H. C., \& Kanbach,
                G. 2002, {\aap}, 391, 225

\bibitem[Tashiro et~al.(1995)]{Tashiro95} Tashiro, M., et~al. 1995,
                {\pasj}, 47, 131

\bibitem[Timmer \& K\"onig(1995)]{Timmer95} Timmer, J., \& K\"onig, M. 1995,
                {\aap}, 300, 70

\bibitem[Uttley, McHardy \& Vaughan(2005)]{Uttley05} Uttley, P., McHardy, I. M., \& Vaughan, S., 2005,
                {\mnras}, 359, 345

\bibitem[Van Cleve \& Caldwell(2010)]{Van10} Van Cleve, J., \& Caldwell,
                D. 2010, Kepler Instrument Handbook, KSCI-10933-001,
                NASA Ames Research Center, Moffett Field, CA
                (http://keplergo.arc.nasa.gov/Instrumentation.shtml)

\bibitem[Villata \& Raiteri(1999)]{Villata99} Villata, M., \& Raiteri,
                C., M. 1999, {\aap}, 347, 30

\bibitem[Wilcoxon(1945)]{Wilcoxon45} Wilcoxon, F. 1945, Biometrics
                Bulletin, 1, 80

\bibitem[Yamada et~al.(2013)]{Yamada13} Yamada, S., Negoro, H., Torii,
                S., Noda, H., Mineshige, S., \& Makishima, K. 2013,
                {\apj}, 767, L34

\end{thebibliography}
\end{document}